\documentclass[twocolumn]{aastex631}

\usepackage{amsmath}
\usepackage{physics}
\usepackage{bbold}
\usepackage{soul}
\usepackage{color}
\usepackage{lipsum}
\newcommand{\revise}[1]{{#1}}
\graphicspath{{./}{revise_figures/}}
\begin{document}

% \title{Cosmic-Ray Feedback from Supernovae in a Stratified Interstellar Medium}
\title{Cosmic-Ray Feedback from Supernovae in a Parker-Unstable Medium }

%% The \author command is the same as before except it now takes an optional
%% argument which is the 16 digit ORCID. The syntax is:
%% \author[xxxx-xxxx-xxxx-xxxx]{Author Name}
%%

\correspondingauthor{Roark Habegger}
\email{rhabegger@wisc.edu}

\author[0000-0003-4776-940X]{Roark Habegger}
\affiliation{University of Wisconsin-Madison Astronomy Department}

\author[0000-0003-4821-713X]{Ellen G. Zweibel}
\affiliation{University of Wisconsin-Madison Astronomy Department}
\affiliation{University of Wisconsin-Madison Physics Department}

\begin{abstract}
\revise{Supernova} energy drives interstellar medium (ISM) turbulence and can help launch galactic winds. What difference does it make if $10\%$ of the energy is initially deposited into cosmic rays? To answer this question and study cosmic-ray feedback, we perform galactic patch simulations of a stratified ISM. We compare two magnetohydrodynamic and cosmic ray (MHD+CR) simulations, which are identical except for how each supernova's energy is injected. In one, $10\%$ of the energy is injected as cosmic-ray energy. In the other case, energy injection is strictly thermal \revise{and kinetic}. \revise{We find that cosmic-ray injections drive a faster, hotter, and more massive outflow long after the injections occur.} Both simulations show the formation of cold clouds (with a total mass fraction $>50\%$) through the Parker instability and thermal instability. \revise{The Parker instability simultaneously produces high mass loading factors $\eta > 10^3$ as it requires few supernovae. We also show how the Parker instability naturally leads to a decorrelation of cosmic-ray pressure and gas density. This decorrelation leads to a significant decrease in the calorimetric fraction for injected cosmic rays, but it depends on having a highly resolved magnetic field.  }
\end{abstract}


\section{Introduction} \label{sec:intro}

\begin{figure*}
    \centering
    \includegraphics[width=\linewidth]{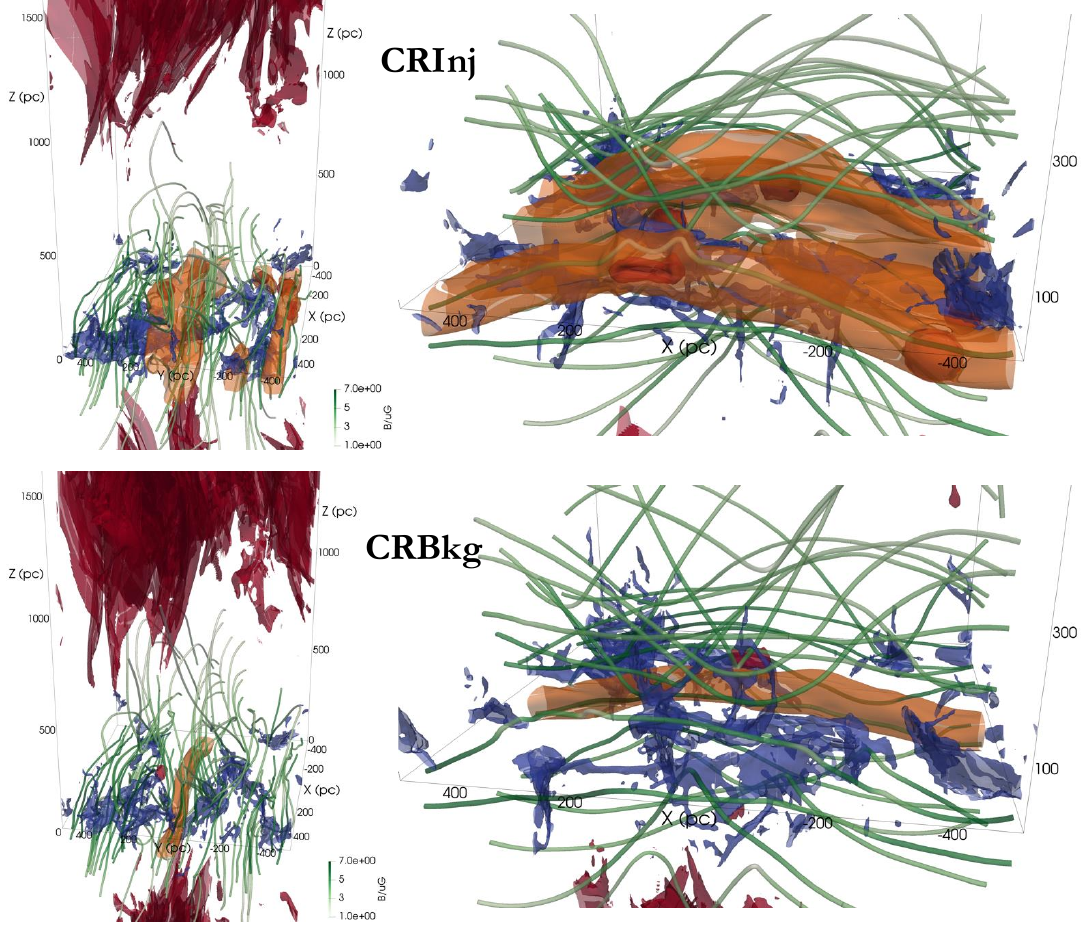}
    \caption{\revise{3D snapshots of each simulation at $t=200\,\mathrm{Myr}$ (see Table \ref{tab:parameters}). The top row shows volume renders from our \texttt{CRInj} simulation, and the bottom row shows renders from our \texttt{CRBkg} simulation. The left side figures show a larger scale picture illustrating the hot wind and exgtended magnetic field, whereas the right side figures show a zoom-in of the midplane region. The blue structures are contours around cold gas $(T \leq 500\,\mathrm{K})$, and the red contours are around hot gas $(T \geq 5 \times 10^4 \,\mathrm{K})$. Green lines trace the magnetic field, and darker green corresponds to stronger magnetic field (see colorbar in left column figures). The orange contours in the $z=0$ surround regions of high cosmic-ray pressure. The red blobs in the midplane show expanding supernova. In the \texttt{CRInj} simulation, these supernovae are encased by orange contours showing where they dumped cosmic ray energy along the local magnetic field.}}
    \label{fig:overview}
\end{figure*}

A tenth, $10\%$, or $0.1$ of anything is a small fraction. Neglecting such a small component is a reasonable assumption in many situations. An example of this small component is cosmic rays in the interstellar medium (ISM): $10\%$ of each supernova's energy is all that is needed to explain the amount of these high-energy particles in our galaxy's ISM \citep{1934Baade,1978Blandford}. More recent work has illustrated that the fraction produced in a shock could be as high as $20\%$ \citep{2014Caprioli}. That fraction could still be considered negligible, especially on the long time scales $(t \gtrsim 100\,\mathrm{Myr})$ and long length scales $(L\gtrsim 1 \,\mathrm{kpc})$ of galaxy evolution, since supernovae drive turbulence and heat gas at small scales \revise{$(L \sim 10 \mathrm{pc})$}.  

However, other work has shown that cosmic rays are not negligible on large scales (see recent reviews \cite{2017Zweibel,2023Ruszkowski,2023Owen}). They are in approximate pressure/energy equipartition with the thermal gas and magnetic field \citep{2001Ferriere}. Gradients in cosmic-ray pressure can drive outflows or fountain flows from a galactic disk, removing gas that could otherwise form stars \citep{2022Chan,2023Tsung,2024Armillotta,2024Thomas}. Bottlenecks created by cosmic-ray streaming can destroy and disrupt cold gas clouds \citep{2017MNRAS.467..646W,2019Wiener,2021Bustard,2022Tsung}. Cosmic-ray transport is faster parallel to the magnetic field, and this anisotropic transport can impact the turbulent energy cascade of the thermal gas \citep{2024Habegger}.

These feedback mechanisms are stifled if cosmic rays are transported out of the galaxy too quickly. In order to keep global galaxy simulations (and cosmological zoom-in simulations) in agreement with observed $\gamma$-ray luminosities, a cosmic-ray diffusion coefficient of about one order of magnitude higher than the standard value of $3 \times 10^{28} \,\mathrm{cm}^2 \,\mathrm{s}^{-1}$ derived from the B/C ratio \citep{2001Jones,2020Evoli} 
is often invoked, following the simulations of \cite{2019Chan}. This diffusion coefficient amplification means that the cosmic rays escape the galaxy before they ever affect the dynamics of the thermal gas. This result brings us back to the idea that cosmic rays produce only a marginal effect; including them just makes stellar feedback slightly more efficient.

In this work, we examine how individual cosmic-ray sources eventually adjust the dynamics and vertical structure of the multiphase ISM, becoming a significant component of stellar feedback. We previously examined how a single localized cosmic-ray energy injection on a magnetic flux tube could disrupt a stratified ISM \citep{2023Habegger}. This work builds on that study to consider \revise{multiple} injections throughout the galactic midplane. Our simulations evolve a slab of stratified ISM, with midplane parameters that match the solar neighborhood. After a significant fountain flow develops in the first $200\,\mathrm{Myr}$, with magnetic field geometry similar to the Parker instability \citep{1966Parker}, the slab reaches a steady state (lasting $\gtrsim 100 \,\mathrm{Myr}$). During the transition to the steady state, the cosmic rays are able to escape quickly without interacting with the thermal gas. \revise{This escape significantly decreases the $\gamma$-ray luminosity.}

\revise{Figure \ref{fig:overview} is a set of volume renders from our simulations at a time of $t=200\,\mathrm{Myr}$}, which illustrate the complex temperature and magnetic field structures that form over time in the stratified, multiphase, and magnetized ISM. The temperature varies over several orders of magnitude \revise{(blue contours surround gas with $T<500\,\mathrm{K}$ and red contours surround gas with $T>5\cdot 10^4\,\mathrm{K}$). Regions of high cosmic-ray pressure (orange contours) follow the magnetic field (in green), inflating magnetic flux tubes. Some magnetic field lines stretch perpendicular to the disk, over $1\,\mathrm{kpc}$ in length, despite initially only being directed horizontally.} Considering the full scale of a galaxy, those magnetic field lines would connect the galactic plane and the galactic halo. We find that these field lines act as escape highways for injected cosmic rays, \revise{taking the simulation below the calorimetric limit}. These extended field lines result from expanding plumes that push the horizontal magnetic field out of the galactic plane. Those plumes drive the creation of cold gas in valleys, the same structure predicted for the Parker instability \citep{1966Parker}. These valleys result in a decorrelation of cosmic-ray pressure and gas density. The decorrelation is visible in Figure \ref{fig:overview}, where the cold gas (blue contours) and high cosmic-ray pressure regions (orange contours) alternate in the direction perpendicular to the mean magnetic field. \revise{While the Parker instability drives this decorrelation in our simulations, the general process of accreting cold gas from the galactic halo should have the same effect as long as the magnetic field is well-resolved.}

In Section \ref{sec:methods}, we detail our simulation setup, numerical methods, and implementation of physical processes.  In Section \ref{sec:results}, we cover the primary results from our simulations. In Section \ref{sec:disc}, we discuss several implications derived from our results and analysis. Finally, in Section \ref{sec:conc}, we provide a short summary and list of key conclusions. 

\section{Methods}\label{sec:methods}
We use the astrophysical magnetohydrodynamic (MHD) simulation code \texttt{Athena++} to evolve a thermal fluid alongside a cosmic-ray fluid \citep{2020Stone,2018Jiang}. The code solves the following system of equations:

\begin{equation}
    \pdv{\rho}{t} + \div{ \left( \rho \vb{u} \right)} = 0
    \label{eqn:cont}
\end{equation}

\begin{equation}
    \pdv{\vb{B}}{t} - \curl{\left( \vb{u} \times \vb{B} \right)} = 0
    \label{eqn:induct}
\end{equation}

\begin{eqnarray}
    \pdv{\rho \vb{u}}{t} + \div{ \left( \rho \vb{u}\vb{u} + \mathbb{1}\left(P_g + \frac{B^2}{2}\right) - \vb{B} \vb{B} \right)} =  \nonumber\\ 
    \rho \vb{g} + \sigma_c \cdot \left(\vb{F}_c - \frac{4}{3}E_c\vb{u} \right)
    \label{eqn:mom}
\end{eqnarray}

\begin{eqnarray}
    \pdv{E}{t} + \div{ \left( \left( E+ P_g + \frac{B^2}{2} \right)\vb{u}  - \left(\vb{B} \cdot \vb{u}\right)\vb{B} \right)} =  \nonumber\\ 
    \rho \vb{g} \vdot \vb{u}+\mathcal{L}(n,T)-\frac{1}{3}\left(\vb{u} + \vb{v}_s\right) \cdot \grad{E_c}
    \label{eqn:e}
\end{eqnarray}

\begin{equation}
    \pdv{E_c}{t} + \div{ \vb{F}_c} = \frac{1}{3}\left(\vb{u} + \vb{v}_s\right) \cdot \grad{E_c}
    \label{eqn:cre}
\end{equation}

\begin{equation}
    \frac{1}{V_m^2}\pdv{\vb{F}_c}{t} + \frac{1}{3}\grad{ E_c} = - \sigma_c \cdot \left(\vb{F}_c - \frac{4}{3}E_c\vb{u} \right).
    \label{eqn:crflux}
\end{equation}

These equations describe the evolution of a magnetized two-fluid system. The first fluid, thermal gas, has a mass density $\rho$, velocity $\vb{u}$, pressure $P_g$. The second fluid, made up of cosmic rays (i.e. a relativistic, non-thermal fluid), is described by a total energy density $E_c$ and flux $\vb{F}_c$. The cosmic-ray transport coefficient $\sigma_c$ is a diagonal matrix which governs how cosmic-ray energy moves in the simulation, $\vb{v}_s$ is the streaming velocity, and the modified speed of light $V_m$ sets the maximum transport speed. We evolve the magnetic field $\vb{B}$ from Faraday's Law in the ideal MHD approximation (Eqn. \ref{eqn:induct}). The remaining variables are the total magnetohydrodynamic energy density of the first fluid $E = P_g /(\gamma-1) + \rho u^2/2 + B^2/2$, the gravitational field $\vb{g}$, and the radiative heating and cooling function $\mathcal{L}(n,T)$, which depends on the number density ($n = \rho /m$, where $m$ is particle mass) and temperature $(T = P_g m/( \rho k_B))$ of the thermal gas. \revise{We assume hydrogen gas with a constant molecular weight of $\mu=1$, so the particle mass is $m = \mu m_H = m_H$.}

For the cosmic-ray transport, we allow the cosmic rays to diffuse along the local magnetic field at the canonical Milky Way value of $\kappa_\parallel = 3 \times 10^{28}\,\mathrm{cm}^2 \mathrm{s}^{-1}$ (\cite{2001Jones}, for more recent \revise{determinations} see \cite{2019Evoli,2020Evoli}). This anisotropic transport is created by splitting the transport coefficient into components perpendicular and parallel to the local magnetic field direction $\hat{b}$ in each cell (see \cite{2018Jiang}): 
\begin{equation}
    \sigma_{c,\mathrm{diff}} = \frac{1}{\kappa_\perp} \mathbb{1} + \left(\frac{1}{\kappa_\parallel} - \frac{1}{\kappa_\perp} \right)\hat{b}\hat{b}.
    \label{eqn:diff_transport}
\end{equation}
For the perpendicular direction, we set the transport rate at $\kappa_\perp= 3 \times 10^{18}\,\mathrm{cm}^2 \mathrm{s}^{-1}$ so there is no diffusion perpendicular to the field over the course of the simulation \citep{2014Desiati}. See \cite{2024Habegger} for an extended discussion on choosing the cosmic-ray diffusion coefficient in this anisotropic, resolved magnetic field treatment.

We also include the effects of cosmic-ray streaming transport. \revise{Streaming at the local Alfven speed is dominant when  cosmic rays generate the same Alfven waves that they gyro-resonantly scatter off \citep{2017Zweibel}. This process leads to additional heating because the Alfven waves dissipate after scattering the cosmic rays, taking energy from the cosmic rays into the thermal gas. The heating of gas (and decrease in cosmic ray energy) are proportional to the streaming speed $\vb{v}_s$ and appear in the energy equations (MHD energy in Equation \ref{eqn:e}, cosmic-ray energy in Equation \ref{eqn:cre}). The additional transport appears in the transport matrix by including the following term:}
\begin{equation}
    \sigma_{c,\mathrm{str}} =
    - \hat{b}\hat{b} \frac{ (\gamma_c -1)\hat{b} \cdot \grad{E_c} }{\gamma_c E_c \vb{v}_s \cdot \hat{b}}
    =-\hat{b}\hat{b} \frac{\hat{b} \cdot \grad{E_c} }{4 E_c \vb{v}_s \cdot \hat{b}}.
    \label{eqn:str_transport}
\end{equation}
Additionally, it is important to note that the streaming velocity is \revise{the Alfven velocity directed down the cosmic ray pressure gradient \citep{1982Mckenzie,1991Breitschwerdt,2017Zweibel,2018Jiang}:}
\begin{equation}
    \vb{v}_s = -\frac{\vb{B}}{\sqrt{4\pi \rho}}
    \frac{\vb{B}\cdot \grad{P_c}}{\left| \vb{B}\cdot \grad{P_c} \right|}.
    \label{eqn:streaming_vel}
\end{equation}

\revise{Note that we use the full gas density $\rho$ in Equation \ref{eqn:streaming_vel}, not the ion density. In reality, the ion density should appear in the definition of the Alfven speed for plasma with an ionization fraction less than one. This simplification only impacts the cold gas which forms in our simulations. }

In cold gas, the ionization fraction decreases so the gas density and ion density disagree. Additionally, Alfven waves are damped making all gyroresonant scattering of cosmic rays weaker. \revise{ Other works have successfully implemented and studied the effect of temperature dependent cosmic ray transport \citep{2018Farber,2021Bustard,2024Armillotta,2024Habegger}.  }

\revise{However, we neglect this effect in these simulations because our chosen diffusion rate is high enough for the cold gas. Diffusion across an $100\mathrm{pc}$ cloud occurs at a speed $\kappa_\parallel/L = 10^3 \mathrm{km}\mathrm{s}^{-1}$. For smaller clouds, the rate of diffusion is even faster. Increasing the rate further by including more complete physics for cosmic ray transport would have a minimal impact on our results because diffusion already rapidly smooths cosmic ray pressure gradients in cold gas.}

The total transport parallel to the magnetic field is the addition of the reciprocals of the transfer terms:
\begin{multline}
    \frac{1}{\hat{b}\cdot \sigma_{c}\cdot \hat{b}} = \frac{1}{\hat{b}\cdot \sigma_{c,\mathrm{diff}}\cdot \hat{b}} +\frac{1}{\hat{b}\cdot \sigma_{c,\mathrm{str}}\cdot \hat{b}} \\ = \kappa_\parallel - \frac{4 E_c \vb{v}_s \cdot \hat{b}}{\hat{b} \cdot \grad{E_c} }.
\end{multline}
This combination gives a final, total transport matrix of 
\begin{equation}
    \sigma_{c} = \frac{1}{\kappa_\perp} \mathbb{1} + \left(\frac{1}{\kappa_\parallel -  \frac{4 E_c \vb{v}_s \cdot \hat{b}}{\hat{b} \cdot \grad{E_c} }} - \frac{1}{\kappa_\perp} \right)\hat{b}\hat{b}.
    \label{eqn:transport}
\end{equation}
This streaming and diffusion implementation is standard, and follows directly from \cite{2018Jiang}. We include it here in an effort to make Equations \ref{eqn:cre} \& \ref{eqn:crflux} clear to readers unfamiliar with \cite{2018Jiang}.

In the simulations used for this work, we \revise{set} $V_m = 0.1 c$. See Appendix C of \cite{2023Habegger} for an extended $V_m$ convergence study. Our largest flow velocities are on the order of $100\,\mathrm{km}\,\mathrm{s}^{-1}$, much lower than our chosen $V_m$.

\begin{table}[]
    \centering
    \begin{tabular}{ccc} 
         Parameter & \texttt{CRInj} & \texttt{CRBkg} \\\toprule
         $T_0/\mathrm{K}$ & \multicolumn{2}{c}{$10^4$} \\\hline
         $\rho_0/ (m_p \,\mathrm{cm}^{-3})$ & \multicolumn{2}{c}{$1$} \\\hline
         $\beta$ & \multicolumn{2}{c}{$1$} \\\hline
         $\beta_\mathrm{cr}$ & \multicolumn{2}{c}{$1$} \\\hline
         \revise{$B_0/ (\mu\mathrm{G})$} & \multicolumn{2}{c}{\revise{$5.89$} }\\\hline
         \revise{$P_c/ (\mathrm{eV}\mathrm{cm}^{-3})$} & \multicolumn{2}{c}{\revise{$0.862$} }\\\hline
         $\Sigma_* / (M_\odot \, \mathrm{pc}^{-2})$ & \multicolumn{2}{c}{$50$}\\\hline
         $H_* / \mathrm{pc} $  & \multicolumn{2}{c}{$100$} \\\hline
         $\kappa_\parallel / (\mathrm{cm}^2 \, \mathrm{s}^{-1})$ & \multicolumn{2}{c}{ $3\cdot 10^{28}$}  \\ \hline
         $\dot{N}_\mathrm{SN} / \mathrm{Myr}^{-1}$ & \multicolumn{2}{c}{ $1$}  \\\hline
         $E_\mathrm{inj,th} / \mathrm{erg} $ & $0.9 \cdot 10^{51}$ & $0.1 \cdot 10^{51}$\\\hline
         $E_\mathrm{inj,cr} / \mathrm{erg} $ & \revise{$1.0 \cdot 10^{51}$} & $0$\\\hline
    \end{tabular}
    \caption{Simulation parameters: initial values of temperature of gas $T_0$, midplane gas density $\rho_0$, plasma beta $\beta$, cosmic-ray beta $\beta_\mathrm{cr}$, stellar surface density $\Sigma_*$ (which sets gravitational acceleration), stellar scale height $H_*$, parallel cosmic-ray diffusion coefficient $\kappa_\parallel$, supernova injection rate $\dot{N}_\mathrm{SN}$, thermal energy per injection $E_\mathrm{inj,th}$, and cosmic-ray energy per injection $E_\mathrm{inj,cr}$}
    \label{tab:parameters}
\end{table}

\subsection{Setup \& Initial Conditions} \label{sec:methods:init}
We present two simulations, with identical initial conditions. The only difference is how supernova energy is injected. In one simulation, named \texttt{CRInj}, each supernova has $10\%$ of its energy injected as cosmic-ray energy instead of thermal energy. \revise{In the second simulation, named $\texttt{CRBkg}$, no cosmic-ray energy is injected with each supernova. Instead, the only cosmic rays are from the initial background we put in to match the observed estimates that cosmic ray pressure is similar in value to the thermal gas pressure (i.e. $P_c \sim P_g$.)} The important parameters for each simulation are shown in Table \ref{tab:parameters}.

The simulations are on a Cartesian grid, with the $\hat{x}$ and $\hat{y}$ directions spanning the midplane of the slab and the $\hat{z}$ direction extending perpendicular to the midplane. The midplane has an area $1\,\mathrm{kpc} \times 1\,\mathrm{kpc}$, but the simulation extends $\pm 2.4\,\mathrm{kpc}$, above and below the midplane, for a total simulation volume of $1\,\mathrm{kpc}^2 \times 4.8 \,\mathrm{kpc}$. \revise{In the $x$ and $y$ directions, the boundary conditions are periodic. In the $z$ direction, we use a vacuum or `diode' boundary condition which only allows outflow and no inflow}.

Each direction has a resolution of \revise{$\Delta x = \Delta y = \Delta z = 5 \,\mathrm{pc}$}, resulting in \revise{$200$} resolution elements in the $\hat{x}$ and $\hat{y}$ directions, but \revise{$960$} in the $\hat{z}$ direction. \revise{This resolution, combined with momentum injection (see section \ref{sec:methods:inject}), is necessary for an accurate model of supernova feedback in a multiphase ISM \citep{2015Kim}. In particular, both are necessary for resolving the Sedov-Taylor phase and the transition to the snowplow phase \citep{2011Draine}.}

The simulations are initially in hydrostatic equilibrium, and the setup is identical to the one in \cite{2023Habegger}. \revise{We use a   gravitational acceleration profile  }
\begin{equation}
    \vb{g}(z) = -2\pi G \Sigma_* \hat{z} \tanh\left( \frac{z}{H_*}\right),
    \label{eqn:gravity}
\end{equation}
which assumes a stellar mass density $\Sigma_*$ with a $\sech^2(z/H_*)$ vertical distribution. \revise{The hyperbolic tangent profile is particularly useful in numerical studies because it is smooth at $z=0$ \citep{1993Giz}.}

Taking the cosmic-ray pressure $P_c$ and magnetic pressure $P_B = B^2/2$ to be proportional to the gas pressure $P_g$, and then assuming the equilibrium is isothermal with $P_g \propto \rho$, the initial condition is
\begin{equation}
    \frac{\rho(z)}{\rho_0} = \frac{P_g(z)}{P_{g,0}} = f(z) = \sech^{\eta }\left( \frac{z}{\eta H} \right),
    \label{eqn:init}
\end{equation}
where $\eta = H_*/H$ is the ratio of stellar scale height $H_*$ to gas scale height 
\begin{equation}
    H = \frac{ k_B T_0}{ 2\pi G \Sigma_* m_p} \left(1+\frac{1}{\beta} + \frac{1}{\beta_\mathrm{cr}} \right).
\end{equation}
Using the function $f(z)$ in Equation \ref{eqn:init}, we can define the cosmic-ray pressure with a cosmic-ray beta $\beta_{\mathrm{cr}} = P_{g,0}/ P_{c,0}$ and the magnetic field with a plasma beta $\beta = 2 P_{g,0} / B_0^2$:
\begin{equation}
    P_c = \frac{P_{g,0}}{\beta_\mathrm{cr}}f(z)
\end{equation}
\begin{equation}
    \vb{B} = \hat{x}\sqrt{\frac{2 P_{g,0}}{\beta}f(z)}.
\end{equation}
All velocities are initially set to zero, and the $\hat{z}$ and $\hat{y}$ components of the magnetic field are set to zero everywhere.

\subsection{Heating and Radiative Cooling} \label{sec:methods:rad}
\begin{figure}
    \centering
    \includegraphics[width=0.86\linewidth]{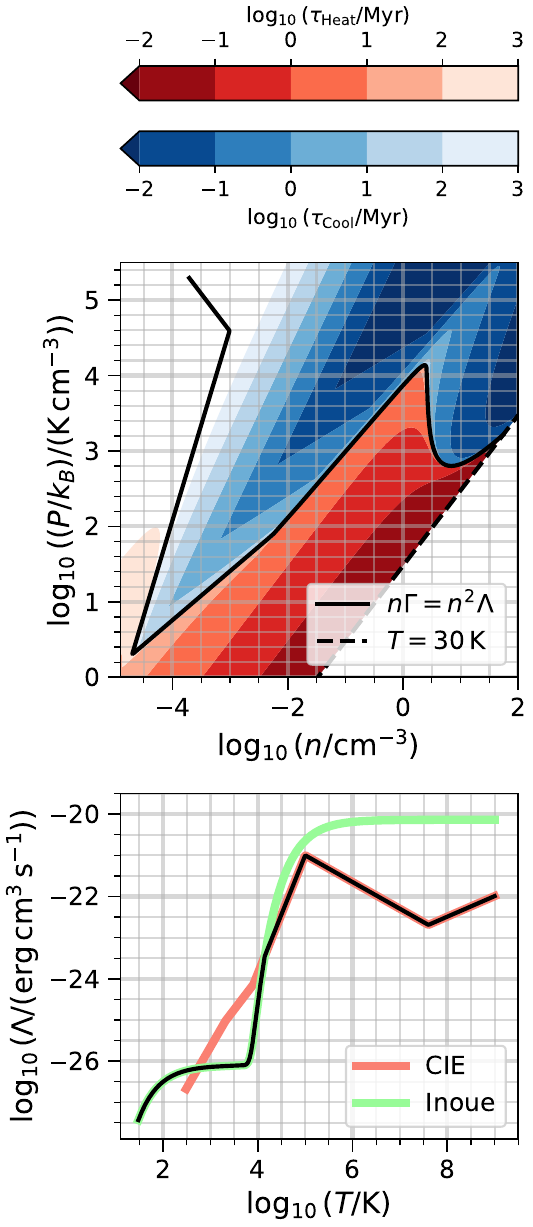}
    \caption{\textit{Top}: Our combined radiative heating and cooling rate $\mathcal{L}$ in Equation \ref{eqn:rad} gives equilibrium states shown as a black line in this gas pressure-gas density phase diagram. We plot our temperature floor as a dashed line in the bottom right of the plot. The contours show the net heating rate in red contours and net cooling rates in blue contours, with darker colors being faster rates. \textit{Bottom}: We use the radiative cooling rate $\Lambda$, as a function of gas temperature $T$, given by the black line. This rate stitches together two functions, one from \cite{2006Inoue} (light green curve) as well as a polynomial fit to the Collisional Ionization Equilibrium (CIE) cooling rate from \cite{1995Rosen}. We transition between these curves at their intersection of $T=1.40413 \cdot 10^{4} \, \mathrm{K}$.}
    \label{fig:cooling}
\end{figure}

We include heating along with optically thin radiative cooling of the ISM in our simulations. We use a source term defined by the net heat-gain function (negative of the heat-loss function):
\begin{equation}
    \mathcal{L}(n,T) = n\Gamma - n^2 \Lambda(T).
    \label{eqn:rad}
\end{equation}
We adopt the cooling function (in units of $\mathrm{erg}\,\mathrm{cm}^{3}\,\mathrm{s}^{-1}$, with temperature $T$ in units of $\mathrm{K}$)
\begin{equation}
    \Lambda(T) = 
    \left\{ 
    \begin{array}{lr}
        \Lambda_\mathrm{IK06}(T) & T < 14041.3  \\
        4.624 \cdot 10^{-36} T^{2.867} &  14041.3 \leq T < 10^5  \\
        1.78 \cdot 10^{-18} T^{-0.65} &  10^5 \leq T < 4 \cdot 10^7   \\
        3.2217 \cdot 10^{-27} T^{0.5} &  4 \cdot 10^7 \leq T 
    \end{array}
    \right.
    \label{eqn:Cooling}
\end{equation}
which combines the cooling function from \cite{1995Rosen} along with the cooling function from \cite{2006Inoue}, which has the functional form
\begin{multline}
    \Lambda_\mathrm{IK06}(T) = 
    7.3\times 10^{21}\exp\left[\frac{-118400}{T+1500} \right] \\ 
    + 7.9\times 10^{27} \exp\left[\frac{-92}{T} 
    \right].
\end{multline}
\revise{These two exponential dependencies are designed to match the expected cooling rates due to Ly$\alpha$ and CII emission \citep{2006Inoue}.}

The cooling functions from \cite{1995Rosen} and \cite{2006Inoue} are plotted in the bottom graph of Figure \ref{fig:cooling}. We stitch them together at $T=1.40413 \cdot 10^4 \,\mathrm{K}$, where they intersect. The resulting curve, our cooling function, is shown as a black line in the bottom figure of Figure \ref{fig:cooling}.

We use the heating rate from \cite{2006Inoue}, a constant $\Gamma = 2 \cdot 10^{-26} \,\mathrm{erg} \,\mathrm{s}^{-1} $ from photoelectric heating by dust grains \citep{2006Inoue,2011Draine}. With all the parameters of Equation \ref{eqn:rad} set, we have an equilibrium curve defined by $\Gamma = n\Lambda(T)$. We show this curve in pressure-density phase space as a solid black line in the top plot of Figure \ref{fig:cooling}. As mentioned in Section \ref{sec:methods:init}, we start with an isothermal atmosphere. The number density ranges from $n=1$ to approximately $n \approx 10^{-4}$. This isothermal atmosphere quickly \revise{$(\lesssim 1 \,\mathrm{Myr})$} relaxes to the equilibrium curve in that density range before the energy injections significantly adjust the structure.

In Figure \ref{fig:cooling}, we also plot contours for the net isochoric cooling and heating time:
\begin{equation}
    \tau = \frac{k_B T}{\gamma-1}\frac{1}{\Gamma - n \Lambda(T)}.
    \label{eqn:tau}
\end{equation}
In Figure \ref{fig:cooling}, the red areas are where $\tau > 0$ in Equation \ref{eqn:tau}, giving net heating. The blue areas are where $\tau < 0$, resulting in net cooling. We plot the isochoric time because we use an operator split approach, including the heating and radiative cooling as an isochoric process. Adiabatic compression and expansion are already included in the energy evolution (see Equation \ref{eqn:e}). This treatment is common, e.g. see \cite{2009Townsend}. Unlike the exact method in \cite{2009Townsend}, we use an explicit integration for the heating and radiative cooling, with order set by the MHD time integrator. Explicit integration is reasonable for our simulations because our cooling time is longer than the simulation's time step \revise{$(\Delta t = \mathrm{CFL} \times \Delta x / V_m \sim 32.6 \,\mathrm{yr})$}, which is set by the modified speed of light $V_m$.

Examining Figure \ref{fig:cooling}, we see at higher densities $n \gtrsim 10^{-2}\,\mathrm{cm}^{-3}$ the gas cools or heats on short timescales, until it again reaches the equilibrium curve where heating and cooling balance. We also plot our floor temperature of $T=30\,\mathrm{K}$ as a dashed line in the top plot of Figure \ref{fig:cooling}. \revise{This floor is necessary to avoid numerical errors, as well as to stop the formation of molecular gas which we do not accurately model.} Whenever the gas reaches the temperature floor, we turn off the radiative heating and cooling terms. If the gas is compressed or expanded by gas motions and brought above the temperature floor, then the terms are again included.

\subsection{Energy Injections}\label{sec:methods:inject}
We inject energy at random locations in the midplane. We restrict the midplane to be cells with vertical coordinate between $z = \pm H_* = \pm 100\,\mathrm{pc}$. \revise{We select a random $x,y,z$ position following that restriction. This random selection neglects the fact that star formation occurs in dense gas \citep{1959ApJ...129..243S}. However, it is a reasonable choice; others have found that injecting supernovae primarily in dense gas can lead to peak-driving which reduces the impact they have when they occur in low-density gas, and thus fails to create a realistic ISM \citep{2015MNRAS.454..238W}.}

The number of injections in each timestep $\Delta t$ is taken from a Poisson distribution, with parameter $\lambda = \dot{N}_\mathrm{SN} \Delta t$ (see Table \ref{tab:parameters}). After calculating the number of injections, we generate their locations using the aforementioned method. This randomization of number of injections makes our simulations stochastic in time, despite having a fixed injection rate $\dot{N}_\mathrm{SN}$. We keep $\dot{N}_\mathrm{SN}$ constant so we can isolate the impact of including cosmic-ray injections. For a more realistic and complete stellar feedback loop, we would need to tie the injection rate to the amount of cold gas in the simulation (assuming the amount of cold gas is a tracer of star formation rate).

\revise{It is important to note that we are considering a low rate of supernovae, $\dot{N}_\mathrm{SN} = 1  \mathrm{Myr}^{-1}$ in comparison to other tallbox simulations. This rate approximately corresponds to a star formation rate of $100 M_\odot  \mathrm{Myr}^{-1}\mathrm{kpc}^{-2} = 10^{-4} M_\odot  \mathrm{yr}^{-1} \mathrm{kpc}^{-2}$. This rate is reasonable in the $\sim \mathrm{kpc}^2$ area near the solar neighborhood, between spiral arms \citep{2022ApJ...941..162E,2025arXiv250617689M}.}

\revise{Once the number of injections and their locations are determined, we inject thermal energy equally in a sphere of radius $20\, \mathrm{pc}$. We also inject kinetic energy (with velocity directed radially outward from the injection location) following the prescription given by Equation 34 in \cite{2015Kim}. Finally, we inject cosmic ray energy in a sphere of radius $50\, \mathrm{pc}$ centered on the same location. Both simulations we present inject the same kinetic energy with each supernova, but we vary the thermal energy and cosmic ray energy injected. In \texttt{CRBkg}, we only inject thermal energy, and in \texttt{CRInj} we inject some cosmic ray energy (see Table \ref{tab:parameters})}.

\subsection{A Note on Parker Instability}\label{sec:methods:Parker}
Our initial conditions are overall similar to setups used to examine the Parker instability \citep{1966Parker}. We tested multiple setups in the single injection simulations and found disruption by the injection \revise{only occurs for setups unstable to the Parker instability} \citep{2023Habegger}. Our initial setup in this work (see Table \ref{tab:parameters}) is linearly unstable to the Parker instability. Since we begin with an isothermal atmosphere, the linear instability criterion (only considering a plane-parallel atmosphere) is \citep{1961Newcomb,2017Zweibel,2018Heintz}:
\begin{equation}
    1 + \frac{1}{\beta}+ \frac{1}{\beta_\mathrm{cr}} > \gamma_g + \gamma_c \frac{1}{\beta_\mathrm{cr}} 
    \label{eqn:lin_inst}
\end{equation}
and plugging in the values from Table \ref{tab:parameters}, \revise{the instability criterion becomes $\gamma_c < \frac{4}{3}$}. Since we include non-advective transport like streaming and diffusion, we expect to satisfy this criteria and the setup is then \revise{unstable to the Parker instability}.

We can also estimate the growth rate of the most unstable mode. Following the dispersion relations derived in \cite{2018Heintz}, we use the 3D Modified Parker (non-zero $\gamma_c$) with streaming transport case. For our setup, we find the most unstable Fourier mode has a wavelength along the magnetic field of \revise{$L_x \sim 0.81\,\mathrm{kpc}$}, and a growth rate of \revise{$\tau \sim 32 \,\mathrm{Myr}$}. Additionally, all the the wavelengths above \revise{$L_x \geq 0.32\,\mathrm{kpc}$} are unstable. \revise{Those large scale wavelengths are all above the characteristic size of a fully evolved supernova supernova remnant, which generally expand to be $100\,\mathrm{pc}$. }

\section{Results} \label{sec:results}

We now describe the results of our two simulations, named \texttt{CRInj} and \texttt{CRBkg}. A key similarity between the simulations is the development of two stages in time. We describe this time evolution before a detailed examination of each stage. The first stage is related to the exponential growth of the Parker instability, while the second stage is a steady state which develops after the Parker instability has disrupted the initial hydrostatic equilibrium. Finally, we compare our simulations to observational diagnostics. 

\begin{figure*}
    \centering
    \includegraphics[width=\linewidth]{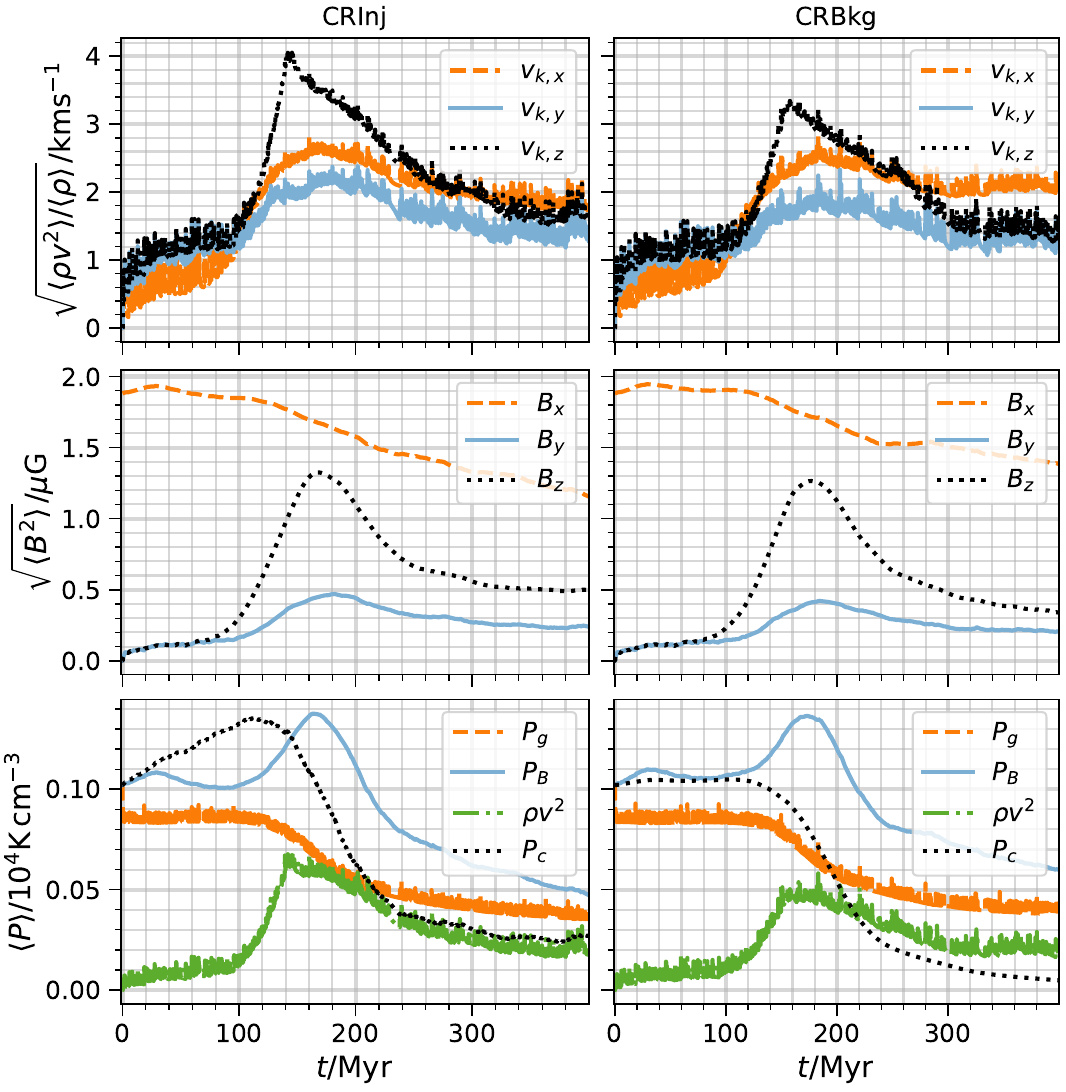}
    \caption{\revise{The time evolution of simulations \texttt{CRInj} (left column) and \texttt{CRBkg} (right column), with each quantity averaged over the entire simulation volume. The top row shows the average velocity divided into its directional components $(\hat{x},\hat{y},\hat{z})$, and the middle row shows the same for magnetic field strength. Compared to the simulation with only an initial background of cosmic rays, additional cosmic ray injections amplify the vertical velocity. The exponential growth of vertical magnetic field is related to the Parker instability (see Section \ref{sec:results:parker}). The bottom row compares thermal, magnetic, turbulent, and cosmic-ray pressures.  The average cosmic ray pressure reaches a steady state near the kinetic pressure and below the average gas and magnetic pressures.  }}
    \label{fig:energy}
\end{figure*}

\begin{figure}
    \centering
    \includegraphics[width=0.95\linewidth]{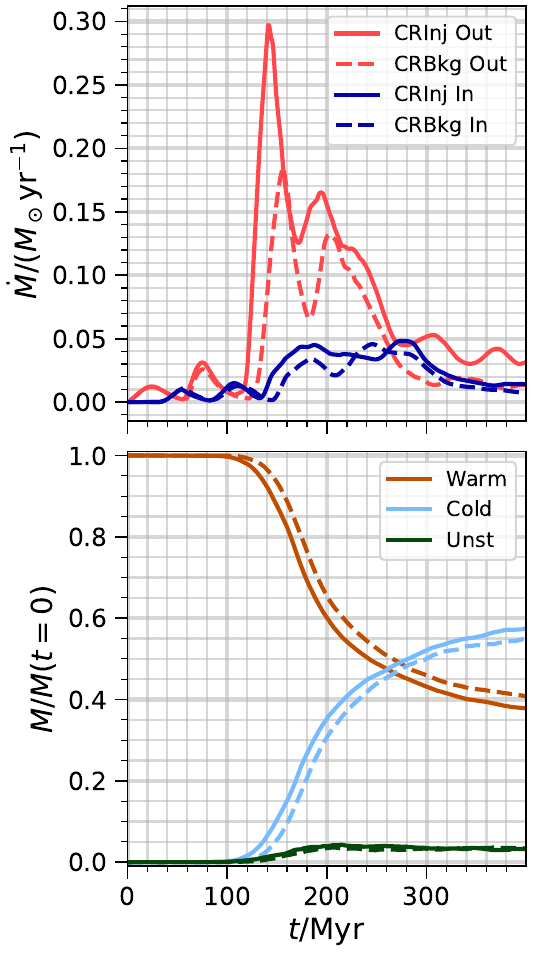}
    \caption{\textit{Top}: Mass outflow and inflow rates for regions above and below the midplane $(|z|>1\,\mathrm{kpc})$. Solid lines show the \texttt{CRInj} simulation and dashed lines show the \texttt{CRBkg} simulation. \revise{The outflows and inflows are out of phase, suggesting the simulations produce a fountain flow. Since the effective star formation rate in our simulation volume is $\sim 10^{-4} M_\sun \mathrm{yr}^{-1} \mathrm{kpc}^{-2}$, the outflows have mass loading factors $\eta = \dot{M}_\mathrm{out}/\dot{M}_\mathrm{SFR}> 10^3$.} \textit{Bottom}: Mass fraction of different phases in terms of the initial mass in the simulation. After $t=0$, the sum of these phases can be less than one because mass can flow out of the top and bottom boundaries. Again, the solid lines show the \texttt{CRInj} simulation and the dashed lines show the \texttt{CRBkg} simulation.}
    \label{fig:Mass}
\end{figure} 

\subsection{Time Evolution} \label{sec:results:time}
We run the simulations for $400\,\mathrm{Myr}$. Initially, the entire setup is destabilized by many energy injections. \revise{These perturbations drive the Parker instability, immediately producing a fountain flow which becomes nonlinear at approximately $100\,\mathrm{Myr}$ into the simulation.} At $200\,\mathrm{Myr}$ the simulation begins to settle into a steady state structure. This steady state lasts from approximately $250\,\mathrm{Myr}$ until the end of the simulation at $400\,\mathrm{Myr}$.

These evolutionary stages are apparent in Figure \ref{fig:energy}, \revise{which illustrates how the average velocity, magnetic field strength, and pressure components in the simulation vary over time.} The left column of Figure \ref{fig:energy} shows the results from the \texttt{CRInj} simulation, and the right column shows the results from the \texttt{CRBkg} simulation.

In the top row of Figure \ref{fig:energy}, we show the \revise{directional components of velocity averaged over the simulation volume.} During the initial destabilization, most of the motion ends up being vertical, perpendicular to the midplane. \revise{Before $t \sim 250\,\mathrm{Myr}$, the \texttt{CRInj} simulation produces more motion in every component direction than the \texttt{CRBkg} simulation, but most significantly in the vertical direction.} Therefore, the additional cosmic-ray pressure from the injections has a significant impact on the dynamics of the gas. In the eventual steady state after $t\sim 250\,\mathrm{Myr}$ the \revise{average velocity in each direction levels off to a constant value.} The \revise{velocity} parallel to the mean magnetic field ($\hat{x}$ direction) ends up being the largest. The spikes are due to individual energy injections. A key difference between the simulations in the steady state is the ordering. In the \texttt{CRInj} simulation, the vertical \revise{motion} is distinctly larger than the planar motion perpendicular to the magnetic field ($\hat{y}$ direction). However, the \texttt{CRBkg} simulation has these \revise{velocities} at about the same value. 

The middle row of Figure \ref{fig:energy} shows the time evolution of the \revise{directional components of magnetic field averaged over the entire simulation volume}. Both simulations have a similar evolution - the magnetic field starts only in the $\hat{x}$ direction, and this energy decays because magnetic flux in the $\hat{x}$ and $\hat{y}$ direction can escape out of the top and bottom of the simulation box. The \revise{magnetic field strength} in the vertical direction increases, exponentially, until approximately $t\sim 180\,\mathrm{Myr}$.  In Section \ref{sec:results:parker}, we show this increase is tied to the Parker instability's characteristic structure. After $t\sim 180\,\mathrm{Myr}$, the \revise{magnetic field} decays. This decay might be offset by amplification due to shearing if we had included galactic rotation in our models. Additionally, we observe numerical magnetic reconnection, which contributes to the \revise{decrease in magnetic field strength}. \revise{Overall, the loss in horizontal magnetic field illustrates how the Parker instability can efficiently decrease, redistribute, and reorient the magnetic field} 

The bottom row of Figure \ref{fig:energy} compares the evolution of thermal, magnetic, turbulent, and cosmic-ray \revise{pressures averaged over the entire simulation volume}. \revise{Initially, the midplane gas, magnetic, and cosmic ray pressures are all $10^4 k_B \mathrm{K} \mathrm{cm}^{-3} $ (see Table \ref{tab:parameters}).} \revise{The total turbulent pressure is initially zero, but it rises to the order of magnitude of the other pressure components.} \revise{The thermal and turbulent pressures have} continual spikes as a result of the injections. \revise{The thermal pressure also has an underlying } decrease as cold, dense gas forms from optically thin radiative cooling. The total magnetic \revise{pressure} reflects the decrease in the $\hat{x}$ directed component and the spike in vertically directed magnetic field. Finally, the cosmic-ray \revise{pressure} shows the biggest difference between the two simulations. The \texttt{CRInj} simulation shows a linear increase early on as the cosmic-ray energy injections add to the background cosmic-ray \revise{pressure}. Both simulations show a decrease in cosmic-ray \revise{pressure} which is tied to the increase in vertical magnetic field. Since the cosmic rays \revise{move along the direction of} the local magnetic field, there needs to be magnetic flux directed out of the top and bottom of the simulation to allow the cosmic-ray energy to escape. As the vertical magnetic field \revise{strength} decreases, the cosmic-ray \revise{pressure} levels off in the \texttt{CRBkg} simulation. The \texttt{CRInj} simulation also levels off, but \revise{to a non-zero value} due to continuing cosmic-ray energy injections.

\begin{figure*}
    \centering
    \includegraphics[width=0.9\linewidth]{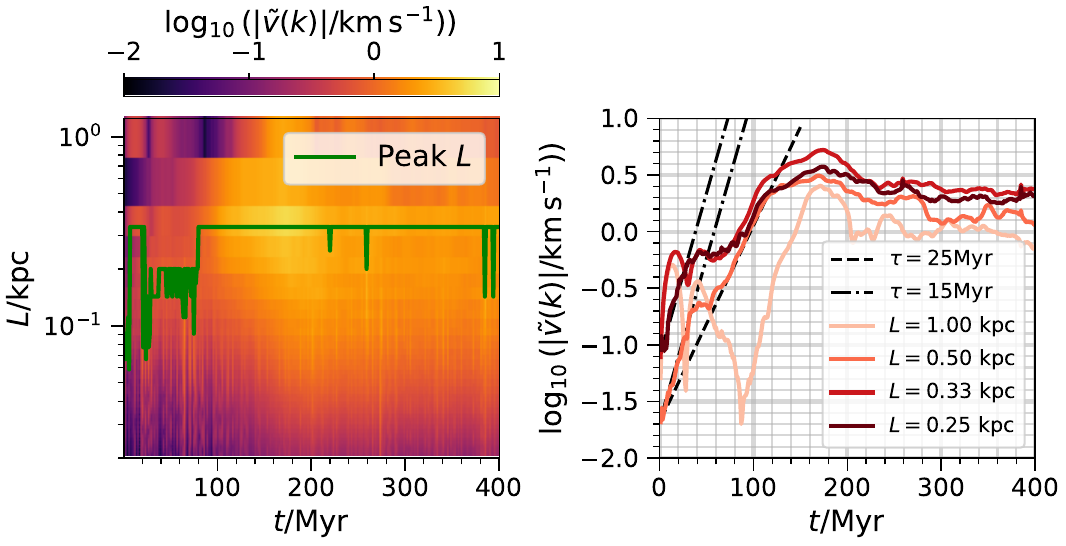}
    \caption{\textit{Left}: Kinetic energy spectrum $\tilde{v}(k)$ (in units of $\mathrm{km} \, \mathrm{s}^{-1}$) for the \texttt{CRInj} simulation at each $1\,\mathrm{Myr}$ time dump. The color map shows the amplitude of velocity fluctuations at each wavelength \revise{$L = 2\pi/k$} and the green line traces the peak wavelength of the spectrum. \revise{The spectrum is primarily driven by supernovae, at a scale of $L\sim20\mathrm{pc}$. Eventually, the larger scale $L\gtrsim 0.33 \mathrm{kpc}$ mode dominates.} \textit{Right}: Amplitude over time of four key Fourier modes, along with \revise{line showing the growth rate of the $L\gtrsim 0.5 \mathrm{kpc}$. The $L = 0.5 \mathrm{kpc}$ mode grows as a result of the Parker instability, independent of the large scale modes until it reaches their level. While that mode never becomes dominant, it feeds the smaller scale modes once it grows to the same strength at $t\sim 80\,\mathrm{Myr}$, causing the $L\gtrsim 0.33 \mathrm{kpc}$ mode to dominate. While the $L\gtrsim 0.5 \mathrm{kpc}$ mode grows consistently until joining the other modes, the small scale modes saturate after $t\sim 20\,\mathrm{Myr}$.}}
    \label{fig:ParkerMode}
\end{figure*}

Moving on from energy evolution, we focus on mass evolution in the simulations. The top graph of Figure \ref{fig:Mass} shows the mass inflow and outflow rates outside of the midplane $(|z| >1\,\mathrm{kpc})$. The solid lines show the \texttt{CRInj} simulation and the dashed lines show the \texttt{CRBkg} simulation. The cosmic-ray injections lead to a larger peak mass outflow (and inflow) rate, and they drive a larger steady outflow at late times $(t>300\,\mathrm{Myr})$. In the \texttt{CRBkg} simulation, the inflow and outflow nearly balance each other at late times. There also appears to be a fountain flow component, where outflows are followed by inflows at a delay of about $\sim 30\,\mathrm{Myr}$. Finally, the two simulation curves are almost exactly the same until $t\sim 100\,\mathrm{Myr}$, \revise{when the Parker instability begins to change the magnetic field orientation}.

In the bottom of Figure \ref{fig:Mass}, we show the mass fractions of different gas phases. \revise{Warm gas has a temperature $4 \cdot 10^3 \mathrm{K}\lesssim T \lesssim 10^5 \mathrm{K}$, whereas unstable gas has a temperature $ 100 \mathrm{K}\lesssim T \lesssim 4 \cdot 10^3 $. Cold gas has a temperature $T \lesssim 100 \mathrm{K}$}. In pressure-density phase space (see Figure \ref{fig:cooling}) the warm gas stretches between the minimum of the equilibrium curve up to the peak above a gas density of $1\,\mathrm{cm}^{-3}$. The unstable gas is between that peak and the minimum just below a density of $10\,\mathrm{cm}^{-3}$, and the cold gas ranges down to the temperature floor of $T=30\mathrm{K}$. It should be noted that the mass fractions in Figure \ref{fig:Mass} do not add to unity at all times. There is a small loss of mass out of the outflow boundaries which causes the total mass to decrease by a factor of $\lesssim 3\%$ over the $400\,\mathrm{Myr}$ of evolution.

Overall, we see that initially $(t\lesssim 120\,\mathrm{Myr})$ some thermally unstable gas is created, which eventually becomes cold gas. The lack of a significant mass fraction of unstable gas is characteristic of weak multiphase turbulence \citep{2024Ho}. That conclusion is consistent with our simulations - most of the kinetic energy in our simulations are vertical, and associated with the bulk outflow in the top panel of Figure \ref{fig:Mass}. \revise{Driving the gas into the strong turbulence regime would require more supernova injections.}

After the unstable gas forms, the cold gas quickly becomes the dominant component of the overall mass fraction after \revise{$t \sim 250\,\mathrm{Myr}$} for both simulations. There is a slight delay in the \texttt{CRBkg} simulation, which is shown with dashed lines. The peak in unstable gas formation tracks with the inflow rate in the top plot of Figure \ref{fig:Mass}. This correlation suggests a connection between the fountain flow and the formation of cold gas via thermal instability. In Section \ref{sec:results:parker}, we show this connection comes from a combination of the Parker instability and Field's thermal instability.

\begin{figure*}
    \centering
    \includegraphics[width=0.9\linewidth]{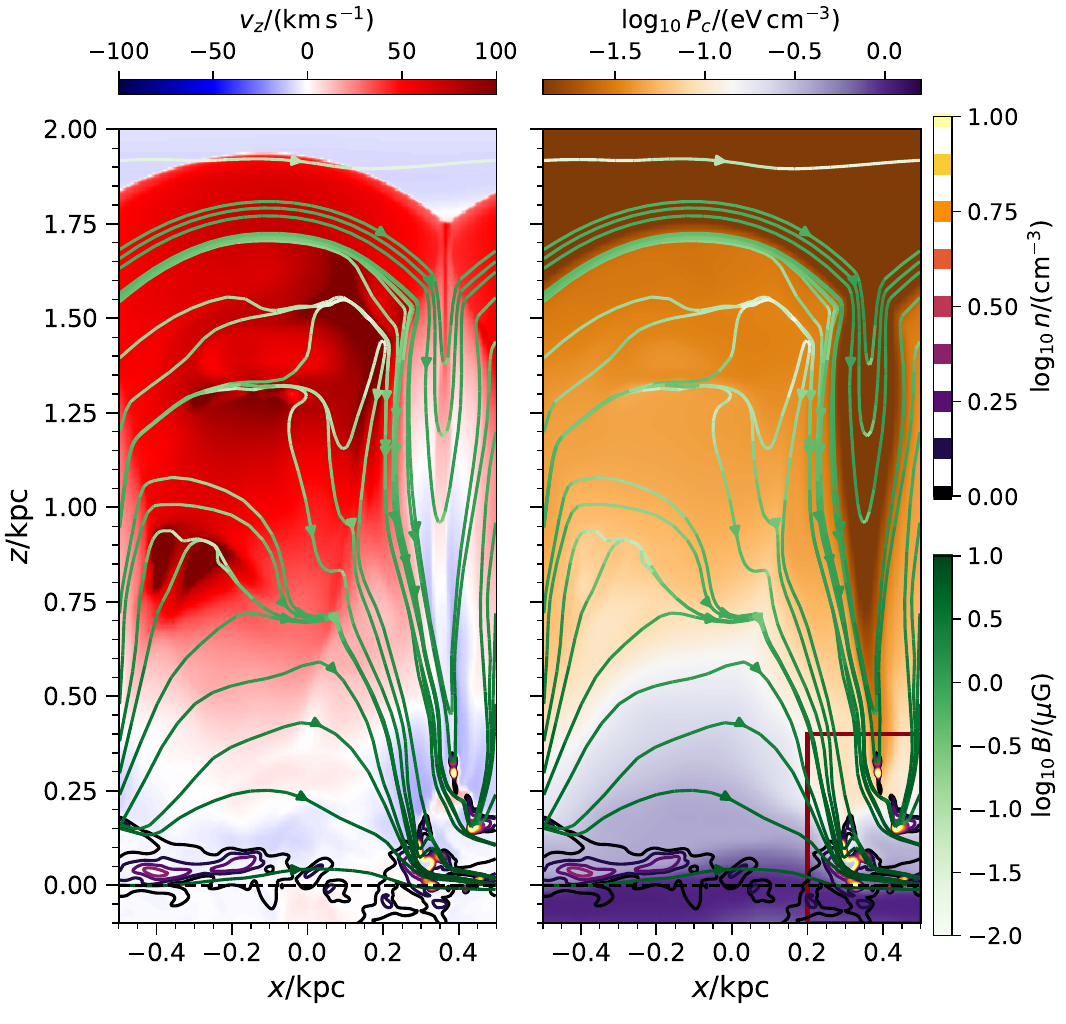}
    \caption{\textit{Left}: Vertical velocity colormap with gas density contours and magnetic field lines overlaid. The large plume centered at $x\sim 0.1 kpc$ is moving upward at a fast rate $(v_z\sim 50 - 100 \,\mathrm{km}\,\mathrm{s}^{-1})$. \textit{Right}: Colormap of cosmic-ray pressure with gas density contours and magnetic field lines overlaid. The density contours highlight the thermally unstable gas phase which forms below where the plume expands into itself (because of periodic boundary conditions). The plume is driven upward partly by the cosmic-ray pressure gradient, which is apparent in the right side plot. Both the left and right simulation slices are taken from the \texttt{CRInj} simulation at \revise{$y=352.5\mathrm{pc}$} and at a time of \revise{$t=135\,\mathrm{Myr}$}. \revise{The crimson box in the right hand figure shows the zoom-in region pictured in Figure \ref{fig:ParkerZoom}}. }
    \label{fig:Parker}
\end{figure*}

\subsection{Parker Instability}\label{sec:results:parker}

Now we examine the initial development of the Parker instability during the beginning of the simulation, $ t \lesssim 200 \,\mathrm{Myr}$. First, we illustrate that the growth of the Parker instability is due to large scale velocity fluctuations which grow linearly out of the small scale energy injections we input. \revise{The primary difference between our simulations and previous simulations is that our perturbations come from discrete supernovae instead of small velocity perturbations (e.g. \cite{2020Heintz}).}

Figure \ref{fig:ParkerMode} shows the evolution of the kinetic energy spectrum. We calculate the spectrum in a $1\,\mathrm{kpc}^3$ volume centered on the midplane - specifically, we only include cells with $|z| \leq 0.5\,\mathrm{kpc}$, matching the length of the $x$ and $y$ directions. On the left side of Figure\,\ref{fig:ParkerMode}, we show the time evolution of the kinetic energy spectrum. Most of the energy is at small scales $L\leq 1/3\,\mathrm{kpc}$. \revise{The wavelength with the most energy at each time is plotted as a green line. For the first $100\,\mathrm{Myr}$, the dominant mode oscillates between $0.1\,\mathrm{kpc}$ and $1/3\,\mathrm{kpc}$. However, the $1/3\,\mathrm{kpc}$ then becomes dominant}. 

\revise{Looking at the right side of Figure \ref{fig:ParkerMode}, it is apparent that the small scale modes quickly saturate after just $20\,\mathrm{Myr}$. Then, the larger scale modes (specifically  $L= 1/2\,\mathrm{kpc}$) eventually rise to join them at $t\sim 100\,\mathrm{Myr}$. results from the growth of the larger scale modes. This large scale mode drives a cascade which amplifies the small scale modes, causing  $L= 1/3\,\mathrm{kpc}$ to become dominant. While the supernova injections are the clear driver of the kinetic energy at early times on small scales  $L\leq 1/3\,\mathrm{kpc}$, it is not obvious what drives the larger scale}.

\revise{The large scale mode $L= 1/2\,\mathrm{kpc}$ grows exponentially at a rate of $\tau=15\mathrm{Myr}$ initially, before saturating at a value lower than the smaller scale modes. While this initial growth could be caused by the fadeaway of supernova explosions, the supernova explosions do not explain the further driving of $L= 1/2\,\mathrm{kpc}$. At $t\sim 60\,\mathrm{Myr}$, the $L= 1/2\,\mathrm{kpc}$ grows exponentially again, at a slower rate of $\tau = 25\mathrm{Myr}$. This rate is closer to the $\tau = 33\mathrm{Myr}$ rate predicted from the linear theory of the Parker instability (see Section \ref{sec:methods:Parker}) than it is to the lifetime of a supernova explosion or to our injection rate}.

\revise{Eventually, even the driving by the Parker instability saturates. This saturation is apparent in the right side of Figure \ref{fig:ParkerMode} for $t\gtrsim 160 \mathrm{Myr}$}. The saturation is significantly faster than $\sim 0.5\,\mathrm{Gyr}$ saturation timescale which previous Parker instability simulations have found \citep{2020Heintz,2023Tharakkal}. \revise{After the saturation point, all the modes decay in amplitude.}

\begin{figure}
    \centering
    \includegraphics[width=1.0\linewidth]{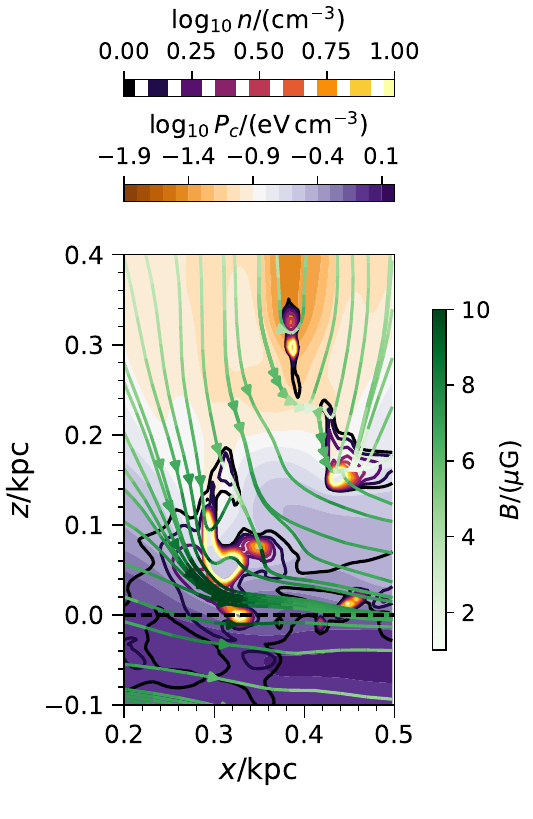}
    \caption{\revise{A zoom-in on dense gas from Figure \ref{fig:Parker}. The condensing gas is in regions of light purple and orange, highlighting that it forms in low cosmic-ray pressure regions. This process decorrelates gas density and cosmic ray pressure. The sinking dense gas also drives an increase in magnetic field strength as it compresses magnetic field lines together. Note that the magnetic field line colors now follow a linear scale as opposed to the logarithmic scale used in Figure \ref{fig:Parker}}.}
    \label{fig:ParkerZoom}
\end{figure}

Although the saturation time is impacted by our limited box size along the magnetic field direction, the rapid outflow begins a little before the saturation (see top plot of Figure \ref{fig:Mass}). This outflow \revise{suggests} the dynamical impact of the Parker instability still happens sooner than \revise{previously expected}. \revise{In fact, it happens at a rate similar to what we saw from single cosmic ray energy injections in a Parker unstable medium \citep{2023Habegger}.}

\begin{figure}
    \centering    \includegraphics[width=0.98\linewidth]{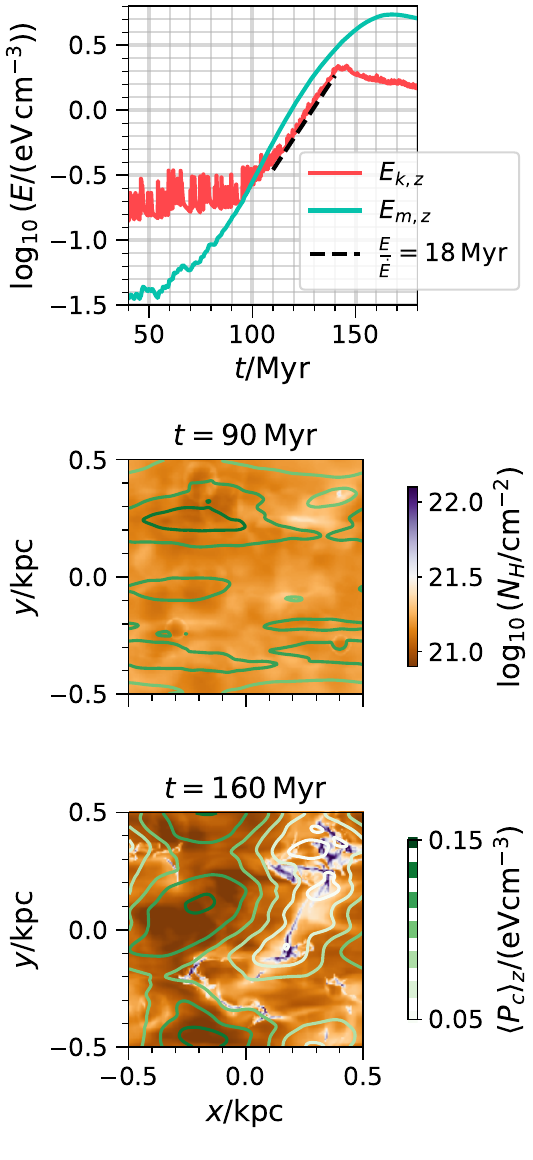}
    \caption{\revise{\textit{Top}: Growth of vertical kinetic and magnetic energy density in the \texttt{CRInj} simulation is exponential with a characteristic time of $\tau \sim 18\,\mathrm{Myr}$. \textit{Middle}: Column density colormap and average cosmic-ray pressure contours at $t=90\,\mathrm{Myr}$. \textit{Bottom}: Same as middle plot, but at $t=140\,\mathrm{Myr}$. The middle panel shows a relatively uniform medium at the time before the magnetic energy density reaches the kinetic energy density. The bottom panel shows the decorrelation of cosmic ray pressure and gas column density after the Parker instability saturates. The cosmic ray pressure is large in the expanding plumes.}}
    \label{fig:ParkerGrowth}
\end{figure}

Now, we focus on the growth and saturation of the large scale modes of the Parker instability. Figure \ref{fig:Parker} shows slices of the \texttt{CRInj} simulation at time \revise{$t=135\,\mathrm{Myr}$} and \revise{$y=352.5\,\mathrm{pc}$}. These slices illustrate saturation of the Parker instability, with plumes which erupt from the simulation's midplane, also producing valleys of cold gas \citep{1966Parker}. The left plot of Figure \ref{fig:Parker} shows the velocity of the gas in the simulation, whereas the right plot shows the cosmic-ray pressure. Both plots have magnetic field lines and gas density contours overlaid. The gas density contours are selected to only be around gas which is thermally unstable.

The plume is moving upward at $v_z \lesssim 100 \,\mathrm{km}\,\mathrm{s}^{-1}$. The cosmic-ray pressure is higher inside of the rising plume of gas, showing that the cosmic-ray pressure gradient is helping to drive the gas flow. The rising plume pushes gas out of the way. \revise{Some gas rolls off the top of the plume and falls back to the midplane along magnetic field lines.} This gas compresses at $z\lesssim 0.25\,\mathrm{kpc}$, forming thermally unstable gas. \revise{The Parker instability picture is generally focused on the formation of these cold gas valleys. However, the roll off of magnetic flux tubes (and gas) \textit{perpendicular} to the mean magnetic field greatly facilitates the rise of the buoyant flux tubes, highlighting the importance of the third dimension \citep{2023Habegger}. I.e. a two-dimensional simulation would have resulted in lower values of $v_z$ since the expanding flux tube would need to lift all gas directly above it \citep{2023Habegger}}.

\revise{Looking at the right plot in Figure \ref{fig:Parker}, it is also apparent that the unstable gas is sinking into gas with a higher cosmic ray pressure. We present a zoom in on this region in Figure \ref{fig:ParkerZoom}.  Figure \ref{fig:ParkerZoom} shows the individual clouds have lower cosmic ray pressures, with colors of light purple, white and orange, whereas the midplane is a dark purple. This results in a decorrelation of cosmic ray pressure and gas density. Additionally, between the gas clouds, there is a magnetic field aligned cosmic ray pressure gradient, allowing cosmic ray pressure to leave the midplane. As the gas sinks, it also compresses the magnetic field, leading to an amplification at the edge of the largest cloud in Figure \ref{fig:ParkerZoom}. }

% \revise{Considering Figure \ref{fig:ParkerZoom} and Figure \ref{fig:energy}, the decorrelation even more apparent on the large scale amplified. Regions of high cosmic ray pressure drive the outflow, but the condensing gas regions will be adjacent to those rising regions}. Therefore, the combination of the Parker instability and thermal instability \revise{decorrelates} cosmic-ray pressure from the local gas density, an important effect in estimating the $\gamma$-ray emission of a galaxy. 

Figure \ref{fig:ParkerGrowth} further illustrates this decorrelation by the Parker instability. During the exponential growth of vertical kinetic and magnetic energy density (top plot), the variations in column density and average cosmic-ray pressure in the $\hat{z}$ direction change significantly \revise{(bottom two plots)}. \revise{Additionally, we find a growth rate of approximately $\tau = \dot{E}/E = 18\,\mathrm{Myr}$ which is faster than the $\tau \sim 33\,\mathrm{Myr}$ growth rate predicted using a linear Parker growth rate.} At \revise{$t=90\,\mathrm{Myr}$} (middle plot of Figure \ref{fig:ParkerGrowth}), the gas is fairly uniform in column density and average cosmic-ray pressure. These quantities are shown with a colormap and contour lines, respectively. 

By \revise{$t=140\,\mathrm{Myr}$} (bottom plot of Figure \ref{fig:ParkerGrowth}), the Parker instability has decreased the column density of the outflowing plume regions and decreased the average cosmic-ray pressure along columns where dense gas is forming. The dense gas regions also correspond \revise{to the footpoint regions} where the magnetic field has become mostly vertical (see right side of Figure \ref{fig:Parker}). This vertical field allows cosmic-ray energy to rapidly escape toward higher values of $z$ \revise{as they stream down the magnetic field aligned pressure gradient}. Eventually that cosmic-ray energy leaves the simulation through the vacuum boundaries at the top and bottom of our simulation. 

The Parker instability and this decorrelation may not remain in a fully turbulent galactic disk, with spiral arms driving additional gas compressions and stirring the magnetic field. But in general, the decorrelation could appear wherever the magnetic field becomes predominantly vertical outside of the midplane. \revise{When vertical field appears, it gives the cosmic rays a pathway to escape and the gas a pathway to sink into the galactic plane and condense}.

\begin{figure}
    \centering
    \includegraphics[width=0.9\linewidth]{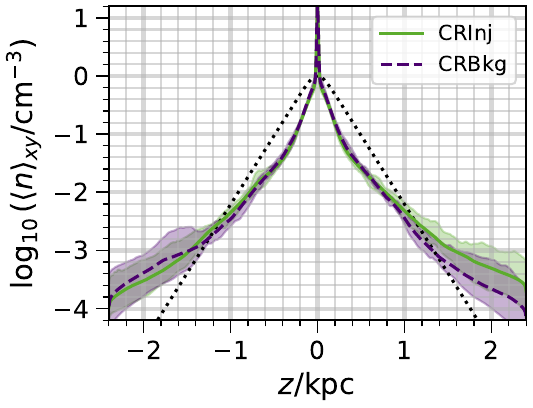}
    \caption{The average vertical density profile over the period \revise{$t=[260,400]\,\mathrm{Myr}$}. The solid green line is from the \texttt{CRInj} simulation and dashed purple line is from the \texttt{CRBkg} simulation. We calculated the profile at every $1\,\mathrm{Myr}$ time dump by averaging over each $xy$-plane and then averaging those vertical profiles over time.  The shaded regions show the full variation (min and max average density value at a given z position) for each simulation. Both simulations form a thin cold gas disk $(n>1 \,\mathrm{cm}^{-3})$, a thin warm disk $(n<1 \,\mathrm{cm}^{-3})$, and a thick warm gas disk $(n<10^{-1} \,\mathrm{cm}^{-3})$.}
    \label{fig:dens_full}
\end{figure}

\begin{figure}
    \centering
    \includegraphics[width=\linewidth]{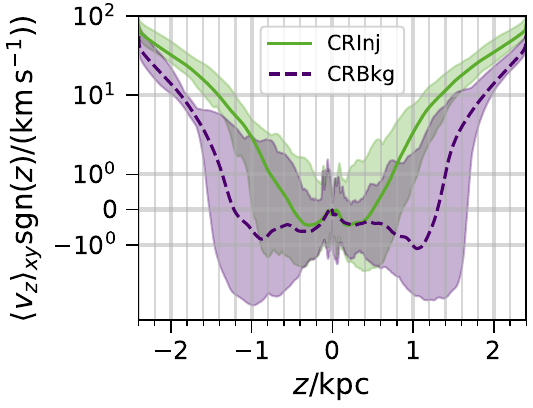}
    \caption{$xy$-plane averaged outflow/inflow velocities from both simulations. The averaged velocity profile over the steady state time frame \revise{$(t\in [260,400]\mathrm{Myr})$} is shown for each simulation. The solid green line is from the \texttt{CRInj} simulation and dashed purple line is from the \texttt{CRBkg} simulation. The shaded regions show the full variation of the profile over the steady state time frame. The \texttt{CRInj} simulation has a \revise{faster} outflow which also starts from closer to the midplane.}
    \label{fig:vel_full}
\end{figure}

\begin{figure*}
    \centering
    \includegraphics[width=0.9\linewidth]{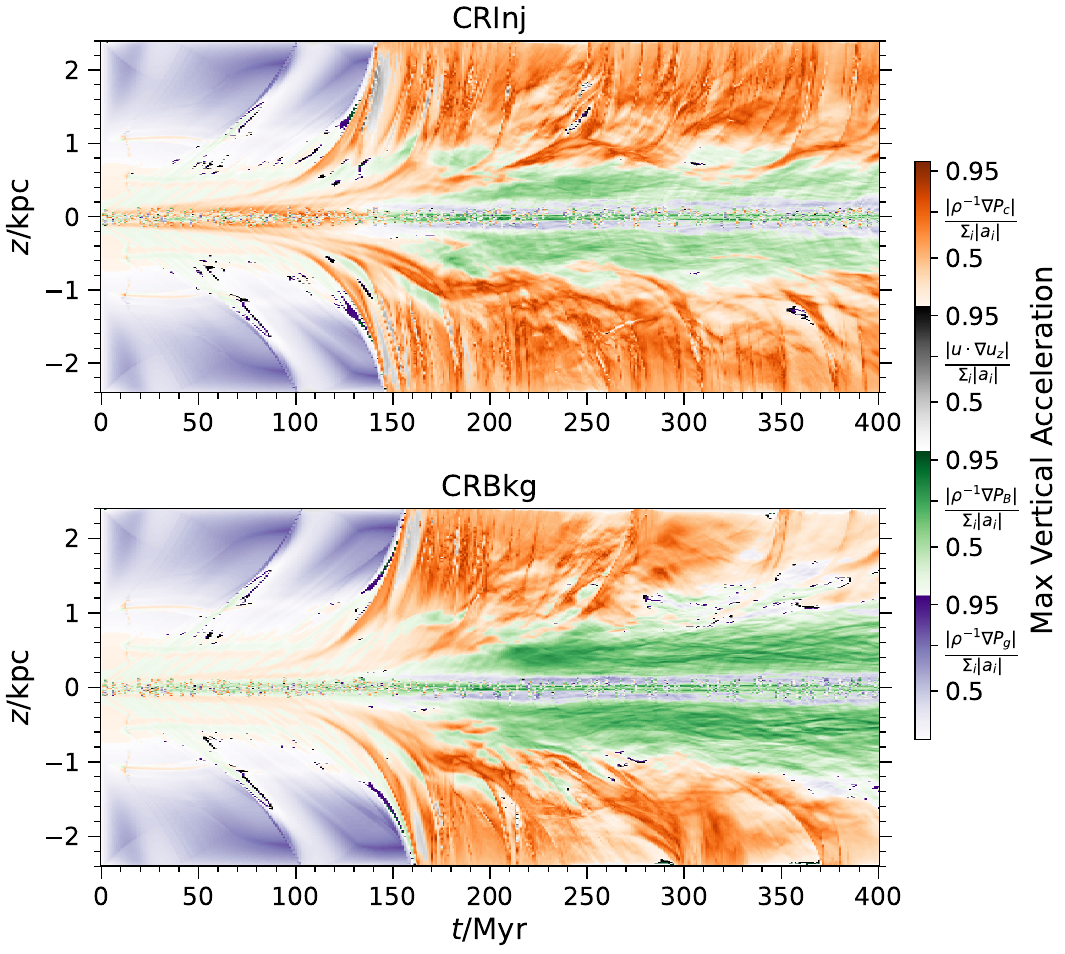}
    \caption{Spacetime diagrams of the dominant acceleration components in the \texttt{CRInj} simulation \revise{(top figure)} and the \texttt{CRBkg} simulation \revise{(bottom figure)}. The lower \revise{quarter} of the colorbar (purple) is for regions where the $xy$-plane averaged vertical \revise{acceleration is predominantly driven by the gas pressure gradient}, the \revise{next quarter} (green) is when the magnetic pressure gradient is dominant, \revise{the following quarter (black) is where the advective motion dominates the acceleration}, and the top \revise{quarter} (orange) is where the cosmic-ray pressure is dominant. The \texttt{CRInj} simulation has much stronger cosmic-ray pressure gradients at later times, once the magnetic field is mainly vertical. }
    \label{fig:comp_pres}
\end{figure*}

\begin{figure*}
    \centering
    \includegraphics[width=0.9\linewidth]{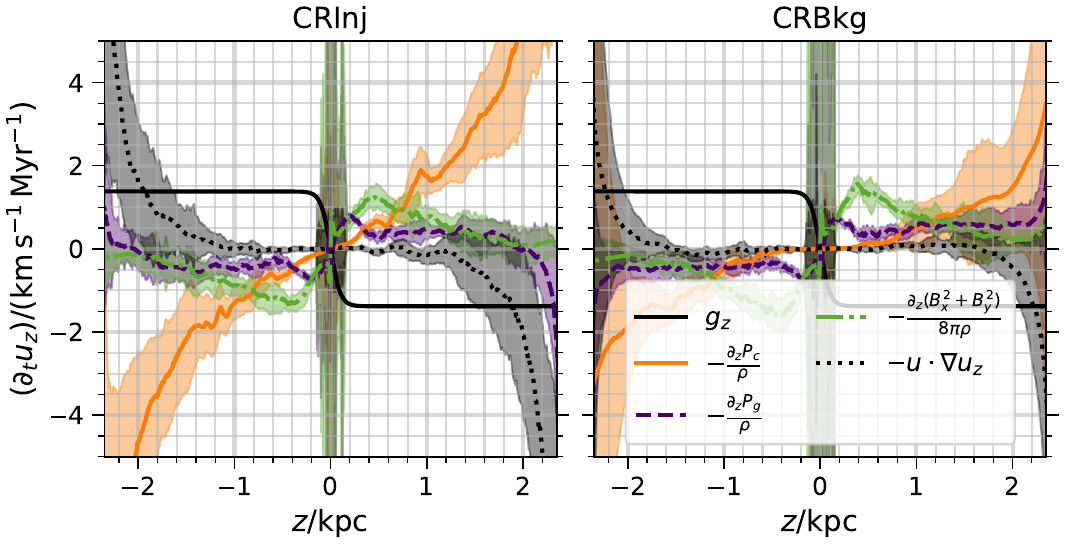}
    \caption{Vertical acceleration by various forces in the \texttt{CRInj} simulation (left plot) and in the \texttt{CRBkg} simulation (right plot). The solid black line shows the gravitational acceleration from the background profile, the solid orange line shows the acceleration from the vertical cosmic-ray pressure gradient, the dashed purple line shows the acceleration from the gas pressure gradient, and the dash-dotted green line shows the vertical magnetic pressure gradient. Each line (except the constant gravitational acceleration) has a shaded region showing the full variation of the acceleration from the given force during the steady state time frame \revise{$(t\in [260,400]\mathrm{Myr})$}.}
    \label{fig:comp_force}
\end{figure*}

\begin{figure}
    \centering
    \includegraphics[width=0.9\linewidth]{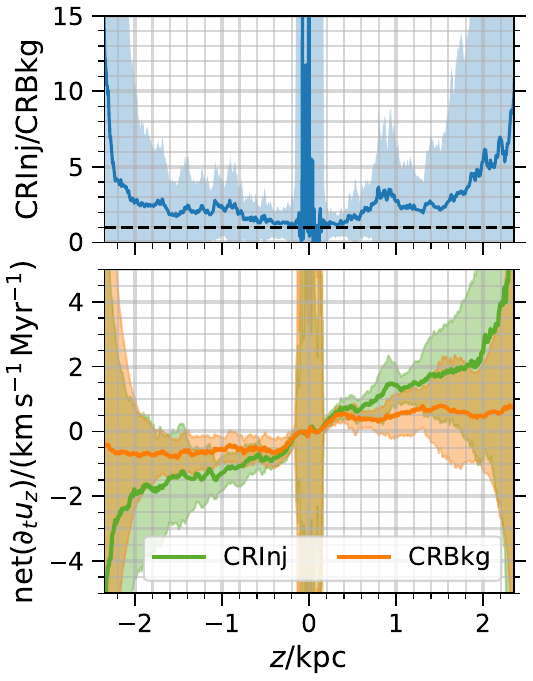}
    \caption{\textit{Bottom}: Net acceleration in each simulation, calculated by adding up the forces in Figure \ref{fig:comp_force}. \textit{Top}: Ratio of net acceleration between the \texttt{CRInj} simulation and the \texttt{CRBkg} simuation. Except for near the midplane, the \texttt{CRInj} simulation always has a larger net acceleration directed away from the midplane (i.e. negative acceleration for $z<0$ and positive acceleration for $z>0$. }
    \label{fig:net_force}
\end{figure}

\subsection{Saturated State} \label{sec:results:steady}

The simulations develop a saturated and steady state after the Parker instability disrupts the original hydrostatic equilibrium. We focus on the time period between \revise{$t=260\,\mathrm{Myr}$} and $t=400\,\mathrm{Myr}$, even though the steady state appears to start as early as \revise{$t=200\,\mathrm{Myr}$} in Figure \ref{fig:energy}. We start the analysis at \revise{$t=260\,\mathrm{Myr}$} to avoid the transition period from the nonlinear evolution of the Parker instability. \revise{This transition is clearer} in Figure \ref{fig:Mass}. There, the outflow and inflow rates, as well as mass fractions of each phase, only level off to consistent values by \revise{$t=260\,\mathrm{Myr}$}. 

In the steady state, we focus on the vertical structure which appears in our simulation. Figure \ref{fig:dens_full} shows average vertical density profiles over time. The density profiles are calculated by averaging over the $xy$-plane at every value of $z$. In Figure \ref{fig:dens_full} the lines show the average profile over the time frame \revise{$t=[260,400]\,\mathrm{Myr}$} for each simulation. The shaded regions show the full variation of the profile over that time frame. The dotted line shows the initial hydrostatic equilibrium. 

Figure \ref{fig:dens_full} shows there is both more gas in the midplane and more gas above $|z| \sim 1\,\mathrm{kpc}$ than at the start of the simulations. The spike in gas density near $z\sim 0$ is cold gas that has settled after forming via thermal instability. There is a switch at large vertical positions, where gas has been pushed out by the Parker instability. This restructuring means a smaller scale height for cold, high density gas and a larger scale height for warm, low density gas. This new quasi-equilibrium appears in both simulations. So, the cosmic-ray injections have a minimal impact on the resultant density stratification, because the averaged profiles are similar in both simulations. The \texttt{CRInj} does end up with slightly denser gas \revise{in the region $z>1\mathrm{kpc}$}, adjusting the scale height slightly (see Section \ref{sec:results:obs}).  

Next, we can look at the flow speed in the saturated state. Figure \ref{fig:vel_full} shows the vertical profile of outflow speed. We calculate the average outflow speed at every value of $z$, integrating over the $xy$-plane. We multiply this value by the sign of the $z$ coordinate to visualize outflow (if the quantity is positive) and inflow (if the quantity is negative). Figure \ref{fig:vel_full} shows the outflow velocity profiles for each simulation averaged over the steady state time frame \revise{$260\,\mathrm{Myr} < t < 400\,\mathrm{Myr}$}, and the shaded regions show the full variation in the profile over that time frame. The \texttt{CRInj} simulation shows a steady outflow beginning at $|z| \sim 0.5 \,\mathrm{kpc}$ which could contribute to a galactic wind. This steady outflow contrasts the state in the \texttt{CRBkg} simulation which shows more fluctuation between outflow and inflow in the time-averaged profile, only having a dominant outflow for \revise{$|z| \geq 1.2 \,\mathrm{kpc}$}.  The fluctuation between outflow and inflow is also apparent in the volume integrated mass outflow and inflow rates plotted in Figure \ref{fig:Mass}.

Presumably, the cosmic-ray injections are driving the outflow in the \texttt{CRInj} simulation, since that is the only difference between it and the \texttt{CRBkg} simulation. To prove this, we calculate the average vertical \revise{acceleration} for gas pressure, cosmic-ray pressure, \revise{magnetic pressure (only the $x$ and $y$ components of the magnetic field contribute), and nonlinear acceleration $\vb{u}\cdot \grad u_z$in every $xy$ plane.} We show the maximum vertical \revise{acceleration} at each vertical position $z$ at each time dump $t$ in Figure \ref{fig:comp_pres}. \revise{The strength of each color shows the fraction of the total acceleration for the dominant component. The notation $\Sigma_i 
|a_i|$ just refers to summing over all the accelerations included in Figure \ref{fig:comp_pres}. Note that we use the absolute value of each acceleration component so that there is no cancellation}. The primary difference between the two simulations is the large cosmic-ray pressure \revise{gradient} supported regions in the steady state period of the \texttt{CRInj} simulation. The cosmic-ray pressure \revise{gradient} is the dominant \revise{acceleration} term outside of the midplane $(|z| \gtrsim 1\,\mathrm{kpc})$ by a large amount, $\geq 95 \%$. This cosmic-ray pressure \revise{gradient} dominance contrasts with the steady state of the \texttt{CRBkg} simulation, \revise{where the cosmic-ray pressure gradient is less dominant, at only $\lesssim 50 \%$ of the acceleration for late times}. 

Both simulations show a pinched region of magnetic support in the midplane, where the gas density is highest, surrounded by regions of gas pressure support. This switch in pressure support within the midplane derives from the switch from cold to warm gas - the cold gas has pressure support from its magnetic field but the warm gas is supported by its own gas pressure. Outside of $|z| \sim 250\,\mathrm{pc}$, the magnetic pressure again takes over in both simulations before falling off. 

Taking a closer look at the \revise{accelerations}, we plot each acceleration component averaged over the saturated time frame in Figure \ref{fig:comp_force}. \revise{The lines show the average acceleration due to the background gravitational field (solid black line) and due to the gradients of cosmic-ray pressure (solid orange line), gas pressure (purple dashed line), magnetic pressure (green dash-dotted line), and nonlinear acceleration (black dotted line).} The shaded regions show the full variation of the acceleration over the saturated time frame. The \texttt{CRInj} simulation has much more acceleration from the cosmic-ray pressure gradient term, which vanishes in the midplane. That is the only difference - the magnetic pressure and gas pressure acceleration profiles in the simulations are similar. They both peak at small $z$ values and then fall off with height. 

The net acceleration in each simulation is shown in the bottom plot of Figure \ref{fig:net_force}. Both simulations have mostly outward (from the midplane) acceleration, which matches with the volume integrated inflow/outflow integration in Figure \ref{fig:Mass}. \revise{These accelerations are only estimates because we do not account for the magnetic tension in the calculation. However, the tension force was predominantly zero at most times because the quantities are averaged over the $xy$ plane at each value of $z$}. The average tension shows up as flat lines when we include them in Figure \ref{fig:comp_force}, so we remove it to make the figure easier to understand. However, the tension does have some variation $(\sim \pm 0.2 \mathrm{km}\,\mathrm{s}^{-1}\,\mathrm{Myr}^{-1})$ which could increase or decrease the net acceleration. 

The top plot of Figure \ref{fig:net_force} shows the ratio of the accelerations, illustrating that the \texttt{CRInj} always has more acceleration. \revise{In some positions it has as much as $10\times$ the acceleration of the \texttt{CRBkg} simulation at large values of $|z|$. In the regions above $0.5\mathrm{kpc}$, the cosmic ray injections provide a factor of $3\times$ acceleration compared to the \texttt{CRBkg} simulation}. This increased acceleration further illustrates the ability of cosmic rays to drive outflows from a galactic disk, even if they do not take over as a forcing mechanism until gas reaches the galactic halo.

\begin{figure}
    \centering
    \includegraphics[width=\linewidth]{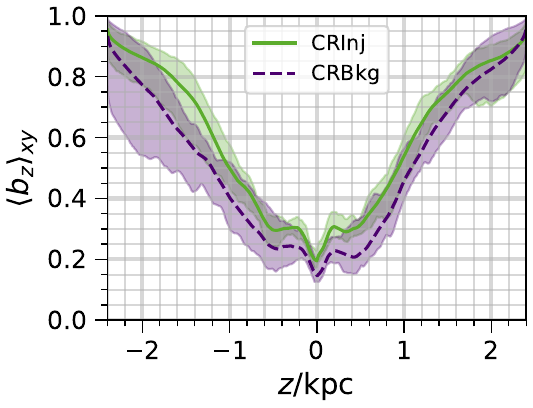}
    \caption{$xy$-plane averaged fraction of the magnetic field directed vertically. The lines show the average vertical magnetic field fraction for the steady state time period \revise{$(t\in [260,400]\mathrm{Myr})$}. The solid green line is from the \texttt{CRInj} simulation and dashed purple line is from the \texttt{CRBkg} simulation. The shaded regions show the full variation of the profile over the steady state time frame. The \texttt{CRInj} simulation produces more vertical magnetic field in the midplane, resulting in a slightly lower height where the vertical field strength becomes dominant.}
    \label{fig:bz_full}
\end{figure}

\begin{figure*}
    \centering
    \includegraphics[width=0.9\linewidth]{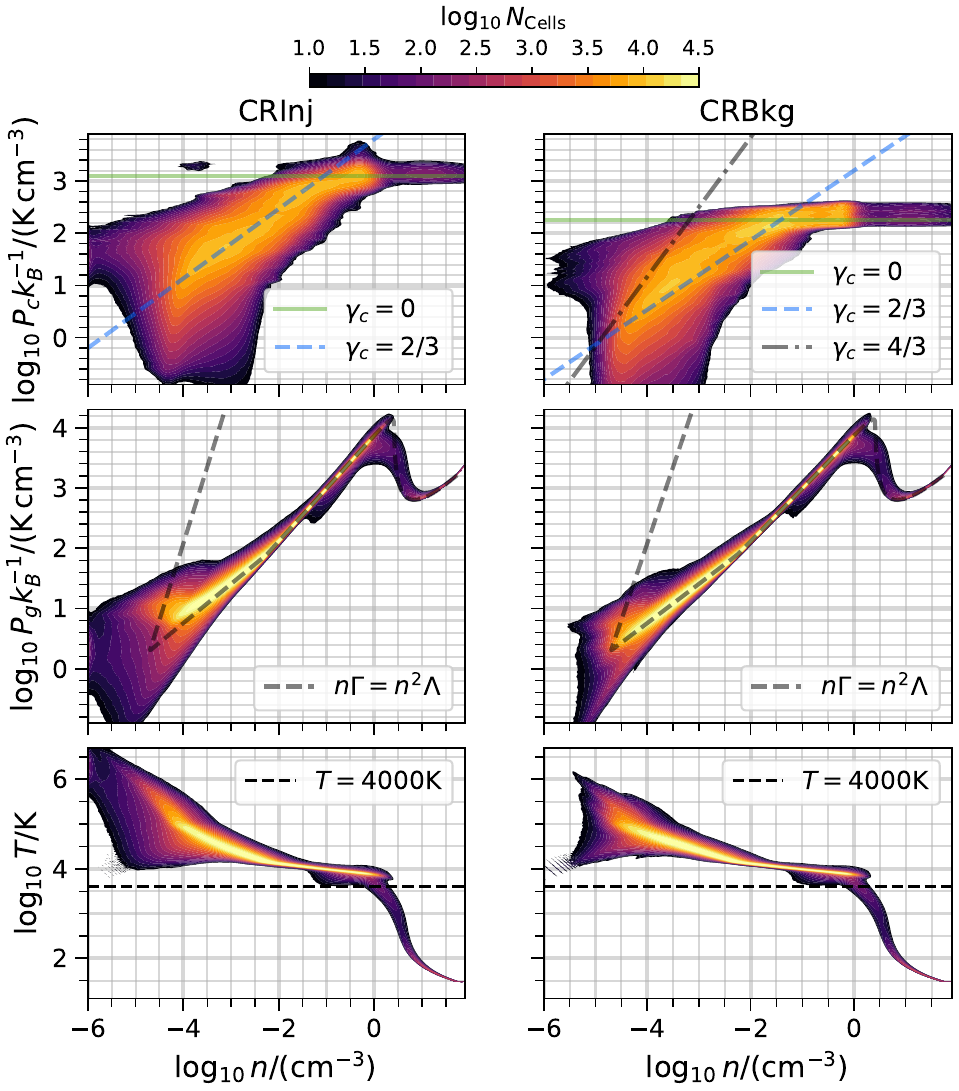}
    \caption{Average phase diagrams for the \texttt{CRInj} simulation (left column) and the \texttt{CRBkg} simulation over the steady state time frame \revise{$(t> 260\,\mathrm{Myr})$}. \textit{Top Row}: cosmic-ray pressure-gas number density phase diagram, which show predominatly diffusive transport in the high density midplane gas and predominantly streaming transport in the diffuse extraplanar gas. \textit{Middle Row}: Gas pressure-gas number density phase diagram. The histogram follows the heating cooling equilibrium curve, shown as a dashed green line (see Figure \ref{fig:cooling}), except the \texttt{CRInj} deviates at high gas density where Alfvenic heating from the cosmic rays pushes the gas off the equilibrium curve. \textit{Bottom Row}: Temperature-gas number density phase diagram. The \texttt{CRInj} ends up with higher temperature gas due to the Alfvenic heating from cosmic-ray streaming. We also plot the cut-off temperature we use to remove cold and thermally unstable gas in Section \ref{sec:results:obs}. }
    \label{fig:phase_diagrams}
\end{figure*}

The dominance of cosmic-ray pressure support in the extraplanar gas deserves more scrutiny. Why does the cosmic-ray pressure support only become significant outside the midplane? To answer this question, we have to examine cosmic-ray transport throughout our simulations.

First, we examine the magnetic field structure, which can impact the cosmic-ray transport. The rapid falling off of magnetic pressure with height in Figures \ref{fig:comp_pres} \& \ref{fig:comp_force} is partly because of a change in magnetic field orientation. The beginnings of that orientation change were apparent in the Parker instability growth pictures in Figure \ref{fig:Parker}. Focusing on the averaged vertical structure, Figure \ref{fig:bz_full} shows the $xy$-plane averaged fraction of the magnetic field in the vertical direction $b_z = |B_z|/B$. We use the absolute value because there is an equal amount of positive and negative vertical magnetic flux when integrating over each $xy$-plane. Above $|z| \sim 1\,\mathrm{kpc}$, the magnetic field becomes predominantly vertical in both simulations during the saturated time frame. There is more vertical magnetic field within the midplane of the \texttt{CRInj} simulation compared to the \texttt{CRBkg} simulation (although it is still subdominant to the horizontal field). As a result, there is less magnetic pressure available to support gas in the \texttt{CRInj} simulation. In the \texttt{CRBkg} simulation, there is more horizontal field in the midplane, but the vertical field still takes over in the extraplanar gas at  $|z| \gtrsim 1\,\mathrm{kpc}$.

Second, we can examine the cosmic-ray transport in the midplane compared to the transport in the extraplanar gas. In the top row of Figure \ref{fig:phase_diagrams}, we show the cosmic-ray pressure-gas number density phase space. The histograms are averaged over the steady state time frame of \revise{$t \in \left[260,400 \right]\,\mathrm{Myr}$}. These figures show a clear differentiation in cosmic-ray transport between the dense gas and diffuse gas. In the dense gas, cosmic rays rapidly diffuse, ending up almost uniformly distributed across orders of magnitude in gas density. Once the cosmic rays escape the dense gas, they enter the extraplanar gas which has a lower density, and high speeds, allowing advective and streaming transport to dominate. \revise{This transition in transport matches what has been seen in previous tallbox simulations with post-processing of cosmic-ray transport, highlighting a transition to advective transport \citep{2021ApJ...922...11A,2024Armillotta}}.

In both our simulations, the streaming transport is dominant over advection. This streaming transport is apparent in the histograms, which do not follow the $\gamma_c  = 4/3$ polytropic law $(P_c \propto n^{\gamma_c})$ for cosmic rays advected with the thermal gas. Instead, it follows a $\gamma_c  = 2/3$ index, which is expected for a steady state system with streaming transport \citep{2019Wiener}. \revise{This result is unexpected given our system is not in a perfect steady state and other transport mechanisms (advection and diffusion) are also included in the simulation}. Likely, since the magnetic field is predominantly vertical and aligned with the cosmic-ray pressure gradient, the streaming becomes dominant. \revise{Additionally, the simulations have a slightly stronger magnetic field than other works, and the turbulence is weak. These changes amplify streaming as a transport mechanism over advection}.

% In the \texttt{CRInj} simulation, there is also a spike just below $\log_{10}(  n / (1\mathrm{cm}^{-3})) =0$. This spike comes from averaging over all the cosmic-ray injections which occur during the steady state time frame. \revise{The cosmic ray injections in the midplane spread cosmic ray energy along the magnetic flux tubes in the midplane. }. 

Figure \ref{fig:phase_diagrams} also shows the gas pressure-gas number density phase space (middle row) and the gas temperature-gas number density phase space (bottom row). The histograms are averaged over the steady state time period. For the most part, the profiles in both simulations follow the thermal equilibrium curve of $\Gamma = n \Lambda$ shown in Figure \ref{fig:cooling} and plotted as a dashed green line in middle row plots. However, the \texttt{CRInj} simulation's histogram departs from this curve at low gas densities, and this low density gas is also at a higher temperature in the temperature-density phase space. This heating comes from the streaming transport of cosmic rays, which takes over in the diffuse gas. At the lower density, the net cooling time increases and is unable to offset cosmic-ray heating, leading to increased temperature. While we have not done any simulations in which streaming is present but cosmic-ray heating is removed from the energy equation, the relative paucity of low density, high temperature in the \texttt{CRBkg} model demonstrates that this is an important effect.

After this examination of the steady state, a clear picture of the interplay between magnetic field orientation and cosmic-ray feedback emerges. The shift to a predominantly vertical magnetic field outside the midplane opens the door \revise{for a cosmic-ray driven outflow}. While injected cosmic rays diffuse in the midplane, they stream and provide a large pressure gradient outside the midplane, producing a faster wind compared to the simulation without cosmic-ray injections. The streaming of cosmic rays also causes them to heat the thermal gas, producing \revise{a hotter outflow}. 

Before we conclude, we test our simulations against observational results. To do this, we calculate and estimate some observable properties of our simulated galactic patch.

\subsection{Observational Comparison}\label{sec:results:obs}

\begin{figure}
    \centering
    \includegraphics[width=0.9\linewidth]{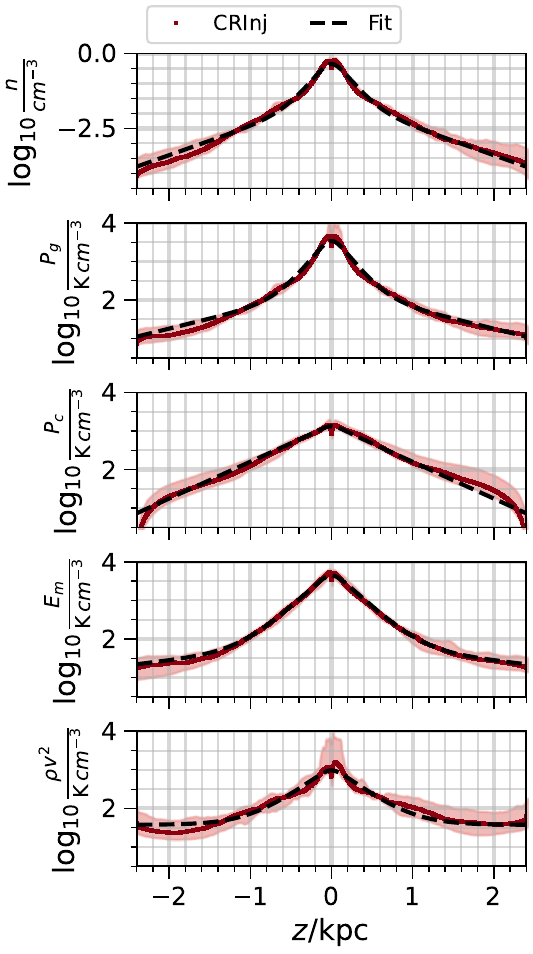}
    \caption{Average vertical profiles from the \texttt{CRInj} simulation, fitted with two independent scale heights. The red shaded regions show the standard deviation of the vertical profiles during the saturated time frame, and the solid red lines show the median profile. The dashed green lines show the initial hydrostatic equilibrium, and the dashed black lines show the best fit to each profile. The profiles are produced by first removing all \revise{cold and unstable gas with $T<3000\,\mathrm{K}$} }
    \label{fig:profiles}
\end{figure}

\begin{figure}[t]
    \centering
    \includegraphics[width=0.9\linewidth]{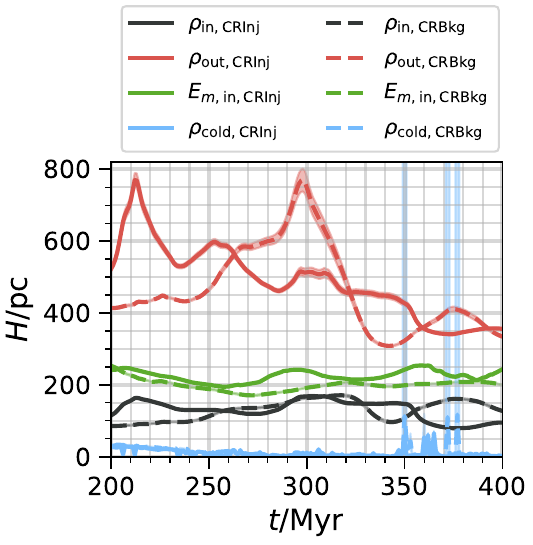}
    \caption{\revise{Gas scale heights in \texttt{CRInj} simulation and \texttt{CRBkg} simulation. Black lines show the inner warm gas density scale heights, and red lines show the outer warm gas density scale heights. The green lines show the scale height for the magnetic energy density, and blue lines show the scale height of the thin, cold disk. Solid lines show \texttt{CRInj} simulation, and dashed lines show the \texttt{CRBkg} simulation.} }
    \label{fig:inflate}
\end{figure}

Using our $xy$-plane averaging process from Section \ref{sec:results:steady}, we can fit vertical profiles and measure scale heights for the gas density, gas pressure, cosmic-ray pressure, and magnetic pressure.

Figure \ref{fig:profiles} shows the average vertical profiles and our median fit to those profiles during the saturated, steady state. The averaged data are plotted within a shaded region, which shows the total variation over the saturated time frame. We removed the cold and thermally unstable gas disk in the midplane \revise{to isolate the impact on warm and hot gas,} by using only gas cells with a temperature $T>4000\,\mathrm{K}$ for the $xy$-plane averaging. We use a linear combination of the base hydrostatic equilibrium profile (see Section \ref{sec:methods:init}):
\begin{equation}
    f ( z | A, H ) = A \left(\sech{ \left[ \frac{z}{H_*} \right]}\right)^{H_*/H}.
    \label{eqn:fitting}
\end{equation}
At each $1\,\mathrm{Myr}$ time dump, this gives us a pair of scale heights $H_\mathrm{inner}$ \& $H_\mathrm{outer}$ and amplitudes $A_\mathrm{inner}$ \& $A_\mathrm{outer}$. We chose the above profile because \revise{it is the equilibrium profile under the influence of our chosen gravitational acceleration (Equation \ref{eqn:gravity})}. The fitted profiles are pictured for gas density and gas pressure in Figure \ref{fig:profiles}, with a central peak near the midplane and large wings outside the midplane.

While the fitted scale heights can be related to an exponential scale height for $z \geq H_*$, \revise{we do not expect them to always match observed scale heights in the Milky Way. This expectation is because the gravitational acceleration profile we use is a simplification of the true acceleration profile}. Instead, we encourage a focus on the structure which our simulations produce: a thinner, dense disk and a thicker disk inflated by non-thermal pressures. A two scale height model matches with the modern understanding of the structure of the Milky Way's thermal gas \citep{2023McClureGriffiths,2024Rybarczyk}. 

For the non-thermal pressure, there is a different story. The cosmic-ray pressure profile in the third row of Figure \ref{fig:profiles} looks significantly different and cannot be fit by the hydrostatic equilibrium profile in Equation \ref{eqn:fitting}. The magnetic energy density profile \revise{in the fourth plot} still has an inner profile and an outer profile, each following Equation \ref{eqn:fitting}. However, the outer profile only takes over at $z\approx \pm 1.5 \,\mathrm{kpc}$ instead of at $z\approx \pm 0.4 \,\mathrm{kpc}$ like the gas density and pressure. \revise{A similar structure also appears for the turbulent pressure in the bottom plot.} Figure \ref{fig:profiles} also illustrates that the gas density, magnetic field, and cosmic rays do not trace each other well. 

In Figure \ref{fig:inflate}, we show the time evolution of scale heights of gas density and magnetic energy density over the saturated state for both simulations. We also include the scale height for the cold gas, which we find by fitting a single component with the form of Equation \ref{eqn:fitting}. We plot these quantities because they are measurable (e.g. gas density via HI emission and absorption, and magnetic energy density through synchrotron radiation). We only plot the inner profile for the magnetic energy density because the values we measure for the outer scale height $(\gtrsim 2\,\mathrm{kpc})$ are untrustworthy. \revise{The outer scale height of the magnetic energy density we find is close to the simulation's vertical box size $(4.8\mathrm{kpc})$, and the outer scale height only takes over above $1.5\mathrm{kpc}$ (see Figure \ref{fig:profiles})}. The only point to be taken away from the outer scale height of the magnetic energy is that it flattens out due predominantly vertical magnetic field at large $z$ (see Figure \ref{fig:bz_full}).

\revise{In Figure \ref{fig:inflate}, it is apparent that the inner scale height of the warm gas, the inner scale height of the magnetic energy density, and the scale height of cold gas, are relatively constant during the steady state time frame \revise{$(t>260\,\mathrm{Myr})$}. The outer scale height of the warm gas varies significantly in both simulations. The inner and outer warm gas scale heights differ by a large factor $(\times 5)$, and the magnetic energy density's inner scale height falls in between the two warm gas components. On average, the \texttt{CRInj} has a slightly larger outer warm gas scale height component and larger magnetic scale height. But the simulations are in overall agreement with each other, and both illustrate the need for two components to fit the warm gas' structure.}

\revise{While the cold gas component initially has a scale height of $\lesssim 50 \,\mathrm{pc}$, it eventually decays to a value equal to the resolution of our simulation $\sim  5 \,\mathrm{pc}$. Without a more complete feedback prescription (i.e. increase the supernovae rate as more cold gas forms), there is nothing stopping the cold gas from falling to $z=0$ (see Section \ref{sec:disc:caveats}) }.

\begin{figure}[t]
    \centering
    \includegraphics[width=0.9\linewidth]{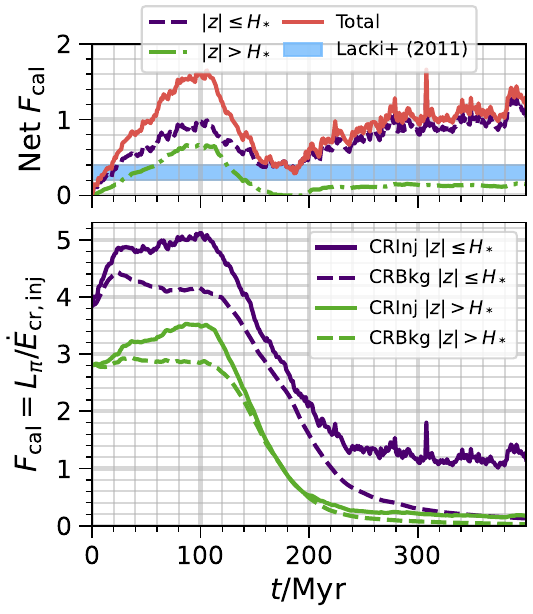}
    \caption{\textit{Bottom}: Fraction of cosmic-ray energy injected which would be lost to hadronic interactions, following the definition of $F_\mathrm{cal}$ in \cite{2011Lacki}. We plot the results for both simulations, with solid lines showing the \texttt{CRInj} simulation and dashed lines showing the \texttt{CRBkg} simulations. We separate the emission from the midplane $|z| \leq H_*$, shown in purple, from the emission outside of the midplane, shown in green. \textit{Top}: We calculate the Net $F_\mathrm{cal}$ by subtracting the \texttt{CRBkg} simulation's $F_\mathrm{cal}$ from the \texttt{CRInj} simulation's $F_\mathrm{cal}$. This subtraction leaves only the $F_\mathrm{cal}$ associated with our cosmic-ray injections. For comparison, the blue shaded region shows observational estimates of $F_\mathrm{cal}$ in starbursting galaxies NGC 253 and M82 \citep{2011Lacki}.}
    \label{fig:FCal}
\end{figure}

\revise{Another observable quantity to consider is $\gamma$-ray luminosity due to hadronic collisions of cosmic rays with thermal particles. These collisions produce a cascade of reactions which includes the emission of $\gamma$-rays.} This luminosity is proportional to the cosmic-ray energy density, providing a useful tool for probing the cosmic-ray content of galaxies other than the Milky Way \citep{2008Guo}. In Figure \ref{fig:FCal}, we show the evolution of the total energy lost in all hadronic interactions for our simulations, \revise{not just the energy emitted in $\gamma$-rays}. We normalize this by the cosmic-ray energy injection rate in the \texttt{CRInj} simulation, following the definition of the $F_\mathrm{cal} = L_\pi / \dot{E}_\mathrm{cr}$ parameter in \cite{2011Lacki}. We calculate the hadronic loss rate $L_\pi$ using the formula for the hadronic loss rate per volume $\Gamma_h$ in Equation 12 of \cite{2008Guo}:

\begin{multline}
    L_\mathrm{\pi} = \int_V d^3x \Gamma_h \\
    = 8.70 \cdot 10^{44}\frac{\mathrm{erg}}{\mathrm{Myr}}
    \sum_\mathrm{cells} \left( \frac{E_c}{\mathrm{eV} \,\mathrm{cm}^{-3}} \right) \left( \frac{n}{\mathrm{cm}^{-3}} \right)
    \label{eqn:Lpi}
\end{multline}
For any subsection of the simulations, we can apply the formula in Equation \ref{eqn:Lpi}.  When calculating $L_\pi$, we partition the each simulation into two regions: the midplane (within a stellar scale height, $|z| \leq H_*$) and everything except the midplane (i.e. cells with $|z| > H_*$).

In the bottom plot of Figure \ref{fig:FCal}, we show $F_\mathrm{cal}$ for both simulations. \revise{The value can be larger than one in our simulations because we do not actually include the loss rate in our simulations. Additionally, these values are not the true measure in the case of the \texttt{CRBkg} simulation which actually has $\dot{E}_\mathrm{cr,inj} = 0$.}

Since most of the gas density is in the midplane and $L_\mathrm{\pi}$ is proportional to gas density (see Equation \ref{eqn:Lpi}), most of the hadronic losses occur in the midplane. Initially, the losses are large because of the initial hydrostatic equilibrium we forced on the system (see Section \ref{sec:methods:init}). After $t \approx 250\,\mathrm{Myr}$, the hadronic losses in the \texttt{CRBkg} simulation vanish \revise{because} much of the initial cosmic-ray background has escaped out of the top and bottom of the simulation. The emission outside the midplane in the \texttt{CRInj} simulation also vanishes. However, in the saturated, steady state we examined in Section \ref{sec:results:steady}, the hadronic losses \revise{level off to a steady value in the midplane of the \texttt{CRInj} simulation. }

To get to a physically realistic and comparable $F_\mathrm{cal}$ parameter, we need to get rid of the initial background cosmic-ray pressure. For an approximation of this, we subtract the \texttt{CRBkg} hadronic loss rate off of the \texttt{CRInj} hadronic loss rate. This net calorimetric fraction is plotted in the top of Figure \ref{fig:FCal} for the midplane, outside the midplane, and their combination. Initially, as cosmic-ray injections are dumped into the simulation volume, the \revise{net} $F_\mathrm{cal}$ increases at a constant rate. Then, at $100\,\mathrm{Myr}$, when the Parker instability begins producing a strong vertical magnetic field, \revise{(see Figures \ref{fig:Parker} and \ref{fig:ParkerGrowth})}, the net $F_\mathrm{cal}$ decreases nearly as rapidly as it rose. With the vertical magnetic field, the cosmic-ray injections are able to escape the simulation faster, producing less hadronic interactions. 

As the simulation saturates, we see a steady growth of the net $F_\mathrm{cal}$ parameter. However, its growth is at a lower rate than early in the simulation, despite our cosmic-ray injection rate being constant (the injections are stochastic in time, but the average rate is constant - see Section \ref{sec:methods:inject}). Since there is still some vertical magnetic fields at these times, there is more escape than at the beginning of the simulation. But that escape is still not enough to agree with observations from \cite{2011Lacki}. The only time our calorimetric fraction gets to the observational values is at the nonlinear saturation of the Parker instability, near \revise{$t\sim 180\,\mathrm{Myr}$ when the vertical magnetic field strength is largest (see Figure \ref{fig:energy})}.

\begin{figure*}
    \centering
    \includegraphics[width=\linewidth]{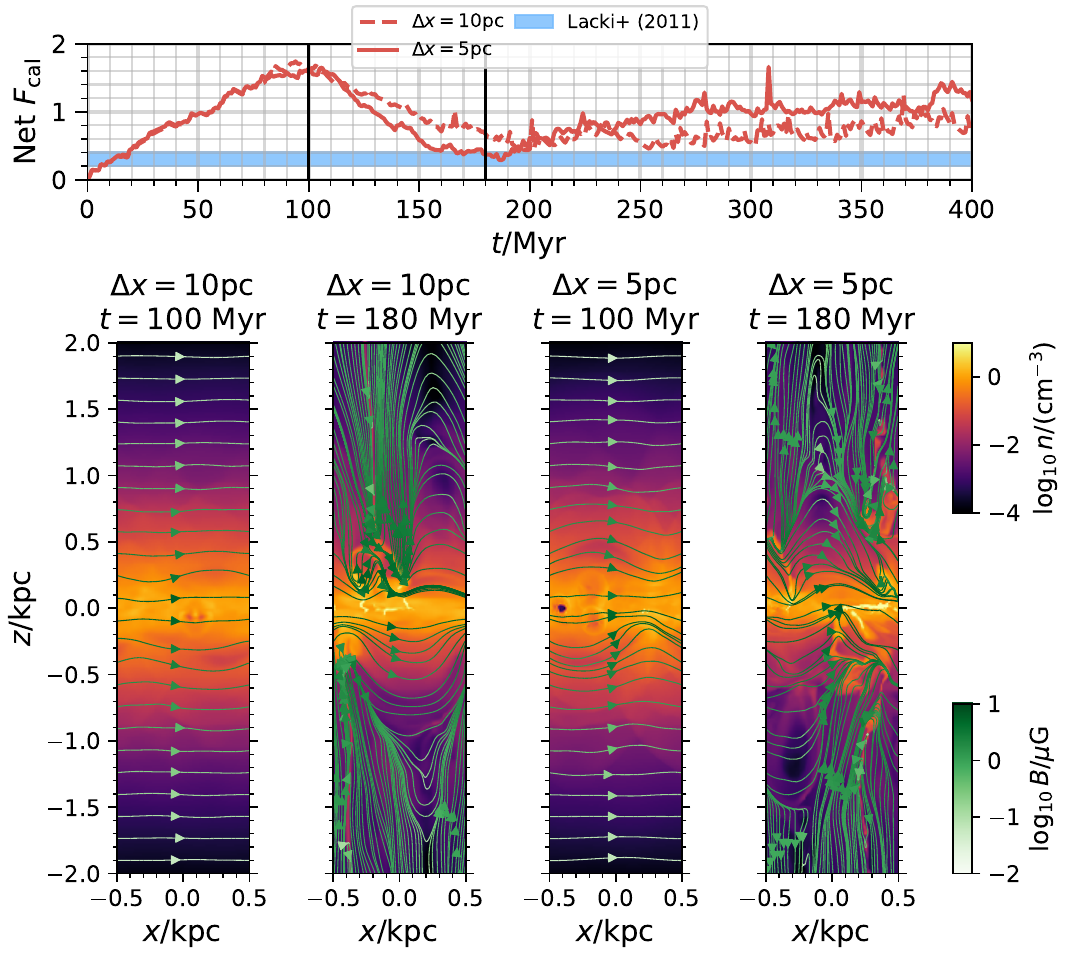}
    \caption{\revise{\textit{Top plot}: Net $F_\mathrm{cal}$ as calculated in Figure \ref{fig:FCal}, in both our low resolution and high resolution simulations. \textit{Bottom plot}: Slices of gas density and magnetic fields in the low resolution and high resolution simulations at $t=100\mathrm{Myr}$ and $t=180\mathrm{Myr}$. The vertical magnetic field facilitates the escape of cosmic rays from the midplane, which decreases the $\gamma$-ray luminosity and the net $F_\mathrm{cal}$.}}
    \label{fig:Resolution}
\end{figure*}

\revise{This minimum point in evolution of $F_\mathrm{cal}$ also depends on resolution. In Figure \ref{fig:Resolution}, we show the evolution of the net calorimetric fraction over time for a $10\,\mathrm{pc}$ resolution simulation and for our fiducial $5\,\mathrm{pc}$ resolution \texttt{CRInj} simulation. Both simulations follow a similar evolution over time, although the higher resolution simulation shows a steeper decrease in net $F_\mathrm{cal}$ during the saturation of the Parker instability. }

\revise{The bottom part of Figure \ref{fig:Resolution} shows how $F_\mathrm{call}$ decreases: the magnetic field becomes vertical between $t=100\,\mathrm{Myr}$ and $t=180\,\mathrm{Myr}$. These field lines allow cosmic rays to escape the simulation, decreasing the $\gamma$-ray luminosity. After the pictured slices, the vertical field lines begin to numerically reconnect. This numerical reconnection is not physical, but it does  disconnect the cosmic rays in the midplane from the outflow boundary conditions in the $\hat{z}$ direction. This disconnect leads to an overestimation of $\gamma$-ray luminosity.  }

\section{Discussion}\label{sec:disc}
\revise{Returning to the original goal of understanding how cosmic ray injections impact stellar feedback, we have seen their primary effect is in accelerating gas at $|z|\gtrsim 0.5\,\mathrm{kpc}$, building the base behind a galactic wind.
This steady outflow lasted for $\gtrsim 100\,\mathrm{Myr}$ and was primarily driven by the cosmic-ray pressure gradient. That gradient was sourced by cosmic rays streaming along vertical magnetic field lines and it was amplified when cosmic ray injections were included. This result is similar to what has been seen in galaxy scale simulations like \cite{2018Farber} and \cite{2024Thomas}, where cosmic rays drive a low density outflow and dense gas is left behind. In fact, our result could form the base of these large scale outflows, or source an outflow along an open magnetic flux tube \citep{1991Breitschwerdt,2022ApJ...926....8S}}.

However, there were also similarities between the two simulations. The cosmic-ray injections had a minimal impact on the vertical density structure (see Figure \ref{fig:dens_full}). \revise{The vertical density profiles of each simulation in the steady state were similar, with the only difference being in the evolution of the scale height of warm gas outside the midplane.}

\revise{Additionally, the Parker instability developed in both simulations and drove a fast outflow while reorienting the magnetic field. The instability launched a short lived and massive outflow before the simulations settle into a steady, but weaker, outflow. The outflow slowed because the free energy (contained in the gravitational stratification and non-thermal pressures) of the initial conditions was released during the nonlinear saturation of the instability. The end effect of the Parker instability in the simulations was the creation of a vertical magnetic field, which facilitated cosmic ray escape and the formation of cold dense gas.}

Our results also suggest some solutions to other questions and problems. The \texttt{CRInj} simulation illustrated that the orientation of the magnetic field \revise{has} a significant impact on cosmic-ray transport. The change in transport has implications for the cosmic-ray calorimeter problem and the vertical stability of the ISM. Before we address these in Sections \ref{sec:disc:cal} \& \ref{sec:disc:feedback}, note that these implications are not isolated to the Parker instability. The change in cosmic-ray transport we see is due to the vertical magnetic field outside of the midplane. This vertical orientation could result from the Parker instability, as it did in our simulations, or from other mechanisms \revise{like superbubble chimneys \citep{2024ApJ...973..136O}}. 

\subsection{$\gamma$-Ray Bright Galaxy Problem} \label{sec:disc:cal}
The steady state structure and change in cosmic-ray transport we see in the \texttt{CRInj} simulation could provide a new solution to the problem of large $\gamma$-ray luminosities in global galaxy simulations. This problem was examined in depth with the FIRE simulations in \cite{2019Chan}. They found that the only way to match the observed $\gamma$-ray luminosities of lower stellar surface density galaxies was to increase the cosmic-ray diffusion coefficient \revise{beyond the observationally estimated value}. 

\revise{The brightness problem can also be couched as a cosmic-ray calorimeter problem.} Observational estimates have shown that smaller (lower $\Sigma_*$) galaxies are not cosmic-ray calorimeters, requiring $F_\mathrm{cal} < 1$ \citep{2011Lacki}. But all of the streaming and Milky-Way diffusion coefficient simulations in \cite{2019Chan} were at or above the calorimetric limit. \revise{Therefore, using the observationally expected transport processes, simulated galaxies end up being cosmic-ray calorimeters, producing too many $\gamma$-rays}. 

We calculated the $F_\mathrm{cal}$ parameter from \cite{2011Lacki} and found that our simulation \revise{reaches} their observational estimates for starbursting galaxies (see top panel of Figure \ref{fig:FCal})\revise{, once we account for our initial cosmic ray background}. 

Our net $F_\mathrm{cal}$ parameter gets close to the observed values without \revise{cosmic ray diffusion amplified above the observed Milky Way value}. Cosmic rays are transported diffusively, and via streaming, along the local magnetic field direction in our simulation, like some of the simulations in \cite{2019Chan}. \revise{Combining} this anisotropic transport with a higher resolution in the diffuse gas allowed the cosmic rays to escape along vertically oriented magnetic field lines. We suspect resolution is the reason this has not been noted before. Since we resolve the entire simulation at \revise{$\Delta x = \Delta y =\Delta z = 5\,\mathrm{pc}$}, we resolve more of the magnetic field structure outside the midplane. For comparison with finite mass codes, a gas mass resolution of \revise{$m_g \approx 0.129 M_\odot$} would be necessary to resolve \revise{$n=10^{-2}\mathrm{cm}^{-3}$} gas at a resolution of \revise{$5\,\mathrm{pc}$}. This mass resolution is much smaller than the $m_g\approx 10^{3}M_\odot$ used in galaxy simulations like \cite{2019Chan}. 

\revise{However, direct calculation of the $F_\mathrm{cal}$ parameter (not the net value) suggests our \texttt{CRInj} simulation still has all cosmic ray energy lost to hadronic collisions (see bottom plot of Figure \ref{fig:FCal}). This coincides with a rise in the net $F_\mathrm{cal}$ }once numerical reconnection occurs, reducing the amount of vertical field. The numerical reconnection occurs where the Parker instability plumes run into each other. While there is still some vertical flux which survives (see Figure \ref{fig:bz_full}), the reconnection decreases the total magnetic energy density (see Figure \ref{fig:energy}). It is necessary to study this process in higher resolution to determine what happens when the effects of numerical reconnection are minimized.

\revise{We suggest spatially resolving the diffuse gas outside the midplane of the galaxy is an important ingredient in solving the problem of simulated galaxies being too bright in $\gamma$-ray luminosity. While the dynamics of diffuse gas outside the midplane} does not require a high resolution, high resolution might be necessary to produce a realistic magnetic field structure. 

Naively, we would expect magnetic field lines should stretch from the midplane, out and up into the galactic halo, and eventually the circumgalactic medium. Without higher resolution in the diffuse gas, these vertical field lines numerically reconnect and the regions become disconnected from the midplane. But if the vertical field lines were still there, connecting the midplane to the diffuse gas, then they would provide an escape highway for cosmic rays. These hidden highways could be the true solution to the calorimeter problem, instead of an increased cosmic-ray diffusion coefficient.

\subsection{Cosmic-Ray Transport and Stability of the Disk-Halo Interface}
\label{sec:disc:feedback}

In our simulations, the cosmic rays are predominantly diffusive at high densities but are transported by streaming at low densities (see top row of Figure \ref{fig:phase_diagrams}). \revise{At lower densities (in completely ionized gas)}, the Alfven speed gets larger, allowing the streaming transport to be larger than diffusive transport. Because of the steady stratification, this density dependence is also a vertical position dependence. This change in transport matches the recent results from other groups who identify a similar switch in cosmic-ray transport outside the midplane \citep{2024Armillotta,2024Thomas}. 

A change in cosmic-ray transport also means a change in feedback. When cosmic rays are streaming at the Alfven speed, they heat the gas (see middle and bottom rows of Figure \ref{fig:phase_diagrams}) by scattering off of Alfven waves they generate themselves. Additionally, the streaming process is an advective process instead of a diffusive process. Instead of smoothing out cosmic-ray pressure gradients like diffusion, any steep cosmic-ray pressure gradients are advected at the Alfven speed. So the switch to streaming means larger cosmic-ray pressure gradients, creating a larger force on the thermal gas, which can help drive outflows (see Figure \ref{fig:comp_pres}). These impacts of streaming have been discussed extensively in other works (e.g. \citealt{2023Tsung}).

The $P_c$-$n$ diagrams in the top row of Figure \ref{fig:phase_diagrams} point to a new idea regarding cosmic-ray feedback at the disk-halo interface. When cosmic-ray transport changes, it also changes the effective polytropic index of cosmic rays. A change in polytropic index changes the stability of the system (note the appearance of $\gamma_c$ in Equation \ref{eqn:lin_inst}). Our simulations show an increase in polytropic index with height. So, the change in cosmic-ray transport with $z$ could naturally \revise{shift the stability of gas parcels moving between the galactic disk and halo}.

We developed this interpretation after reading \cite{2024Hosking}, which illustrated how a plasma's compressibility can impact the structure of metastable, stratified MHD atmospheres. A detailed analysis on the stability of a CR+MHD atmosphere is out of the scope of this paper. \revise{Instead, we just share the idea here and examine it in detail in a forthcoming paper (Habegger, Hosking \& Zweibel, in prep.).}

\subsection{Caveats}\label{sec:disc:caveats}
Before concluding, we would like to detail some caveats to this work and the assumptions we made in designing the simulations. 

\revise{Firstly, the gravitational profile we chose for this equilibrium (Equation \ref{eqn:gravity}) is unrealistic at large $|z|$ outside of the midplane, because (1) it does not include a contribution from the dark matter halo and (2) it assumes an infinite plane of stars. At large $|z|$, the dark matter halo will have more of an impact and the finite size of the galaxy will become important. Overall, these effects will decrease the vertical component of gravitational acceleration at large $|z|$, increasing the outflow rate we find in the simulations, helping to launch a large-scale cosmic-ray driven galactic wind}.

\revise{Next, our radiative heating mechanism of a constant $\Gamma$ for all gas densities and vertical positions is unrealistic. The heating rate should fall off with vertical position $z$, allowing more radiative cooling to occur outside the midplane. This choice makes it harder for gas to cool at large $z$ where gas densities are also lower. However, choosing a constant $\Gamma$ makes the thermodynamic equilibrium curve (top panel of Figure \ref{fig:cooling}) the same for the entire simulation volume, simplifying the analysis}. 

\revise{Also, we do not include cosmic-ray momentum boost of supernovae explosions \citep{2018Diesing}, or a local decrease in the cosmic-ray diffusion coefficient in star formation regions \citep{2021Semenov}. These additional complexities in cosmic-ray transport would primarily change cosmic-ray feedback in the midplane, but would have a smaller impact on larger length scales $(L\gtrsim 1\,\mathrm{kpc})$ and longer timescales $(t > 1 \,\mathrm{Myr})$ we consider}. 

\revise{Finally, our work does not complete the full stellar feedback loop to show how cosmic-ray injections impact the star formation rate. To complete that loop, the injection rate would at least need to be determined by star formation rates (e.g. like in the galaxy scale simulations of \citealt{2019Chan} \revise{or other tallbox simulations like \citealt{2021MNRAS.504.1039R,2023ApJ...946....3K}}). Because we \revise{do} not complete the loop, the physical process we highlight of vertical magnetic field lines being cosmic ray escape highways is just a physical explanation of decreased $\gamma$-ray luminosity. The overall effect on star formation rate of magnetic field lines connecting the disk and halo is still unclear}. 

\revise{However, we suspect the true picture of cosmic ray feedback on star formation rates will require the increased magnetic field resolution. For example, reconsider Figure \ref{fig:Parker}. In cases where the Parker instability is a dominant dynamical process, the decorrelation of cosmic rays and gas density could significantly impact start formation rates. However, that decorrelation was caused by field lines being vertical and not undergoing numerical reconnection. The vertical field lines also allowed cosmic rays to escape the midplane before they could interact with the cold gas, decreasing the $\gamma$-ray luminosity. So, the fundamental mechanism is the production of a vertical field, connecting our results to other setups which may not undergo the Parker instability but have a different way to produce vertical magnetic field}.

\section{Conclusion}\label{sec:conc}
The simulations we have run, \texttt{CRInj} and \texttt{CRBkg}, have illustrated the importance of cosmic-ray injections in driving outflows and structuring the ISM. Here, we list some key results from this study of cosmic-ray feedback:
\begin{enumerate}
    \item The Parker instability drives overturning of gas which turns our initial hydrostatic equilibrium into a two-scale-height vertical gas density structure, \revise{in agreement with observations of the Milky Way.} This overturning and saturation happens on a shorter timescale than previously predicted: $\sim 150 \mathrm{Myr}$ instead of $\sim 500 \mathrm{Myr}$ \citep{2020Heintz}). This faster rate is the result of supernovae injection.
    \item This overturning of gas combined with thermal instability drives the production of thermally unstable gas above the midplane, which then sinks to $z\sim 0$ while cooling. This process occurs between outflowing plumes driven, in part, by cosmic-ray pressure gradients. Our simulations show this process decorrelates cosmic-ray pressure and gas density.
    \item \revise{During the nonlinear saturation of the Parker instability, a more vertically oriented magnetic field forms just outside the midplane}. This magnetic field structure allows for more efficient cosmic-ray escape alongside significant cosmic-ray feedback and a decrease in the calorimetric fraction $F_\mathrm{cal}$.
    \item \revise{The same vertical magnetic field structure also allows for a cosmic ray pressure gradient driven outflow to survive for a long time period $(>100\,\mathrm{Myr})$. This outflow could form the base of a large scale galactic wind, stretching down to $|z| \sim 500\,\mathrm{pc}$.} 
    \item  \revise{The outflow is amplified in both the Parker instability regime and in the steady state that forms afterwards by the individual cosmic-ray injections included in the \texttt{CRInj} simulation.}
    \item \revise{We found that resolution was a significant factor in decreasing the calorimetric fraction $F_\mathrm{cal}$. Numerical reconnection of magnetic field lines disconnects the midplane from the high values of $|z|$, trapping cosmic rays in the midplane. Without high resolution of accurate subgrid modeling of vertical magnetic field, simulations will overpredict $\gamma$-ray luminosity of galaxies}.
\end{enumerate}

\begin{acknowledgments}

\revise{We would like to thank the anonymous reviewer, whose comments significantly improved the paper. In particular, their advice on supernova momentum feedback significantly improved our simulations}.

The authors would like to thank Sophie Aerdker, Bob Benjamin, Ryan Farber, Karol Fulat, Ka Wai Ho, David Hosking, Francisco Ley, Lukas Merten, Peng Oh, Nickolas Pingel, Mohan Richter-Addo, Mateusz Ruszkowski, Aaron Tran, Bindesh Tripathi, and Ka Ho Yuen for their insight and helpful conversations which improved this work. The authors greatly appreciate funding from  funding from NASA FINESST grant No. 80NSSC22K1749 and NSF grant AST-2007323 which supported this work. Resources supporting this work were provided by the NASA High-End Computing (HEC) Program through the NASA Advanced Supercomputing (NAS) Division at Ames Research Center.
\end{acknowledgments}
\software{Athena++ \citep{2020Stone,2018Jiang}, MatPlotLib \citep{2007Matplotlib}, NumPy \citep{2011NumPy,2020NumPy}, AstroPy \citep{2013AstroPy,2018Astropy}, Wolfram Mathematica \citep{Mathematica}}

\vspace{5mm}

% \pagebreak


% \appendix

\bibliography{sample631}{}

\begin{thebibliography}{}
\expandafter\ifx\csname natexlab\endcsname\relax\def\natexlab#1{#1}\fi
\providecommand{\url}[1]{\href{#1}{#1}}
\providecommand{\dodoi}[1]{doi:~\href{http://doi.org/#1}{\nolinkurl{#1}}}
\providecommand{\doeprint}[1]{\href{http://ascl.net/#1}{\nolinkurl{http://ascl.net/#1}}}
\providecommand{\doarXiv}[1]{\href{https://arxiv.org/abs/#1}{\nolinkurl{https://arxiv.org/abs/#1}}}

\bibitem[{{Armillotta} {et~al.}(2021){Armillotta}, {Ostriker}, \&
  {Jiang}}]{2021ApJ...922...11A}
{Armillotta}, L., {Ostriker}, E.~C., \& {Jiang}, Y.-F. 2021, \apj, 922, 11,
  \dodoi{10.3847/1538-4357/ac1db2}

\bibitem[{{Armillotta} {et~al.}(2024){Armillotta}, {Ostriker}, {Kim}, \&
  {Jiang}}]{2024Armillotta}
{Armillotta}, L., {Ostriker}, E.~C., {Kim}, C.-G., \& {Jiang}, Y.-F. 2024,
  \apj, 964, 99, \dodoi{10.3847/1538-4357/ad1e5c}

\bibitem[{{Astropy Collaboration} {et~al.}(2013){Astropy Collaboration},
  {Robitaille}, {Tollerud}, {Greenfield}, {Droettboom}, {Bray}, {Aldcroft},
  {Davis}, {Ginsburg}, {Price-Whelan}, {Kerzendorf}, {Conley}, {Crighton},
  {Barbary}, {Muna}, {Ferguson}, {Grollier}, {Parikh}, {Nair}, {Unther},
  {Deil}, {Woillez}, {Conseil}, {Kramer}, {Turner}, {Singer}, {Fox}, {Weaver},
  {Zabalza}, {Edwards}, {Azalee Bostroem}, {Burke}, {Casey}, {Crawford},
  {Dencheva}, {Ely}, {Jenness}, {Labrie}, {Lim}, {Pierfederici}, {Pontzen},
  {Ptak}, {Refsdal}, {Servillat}, \& {Streicher}}]{2013AstroPy}
{Astropy Collaboration}, {Robitaille}, T.~P., {Tollerud}, E.~J., {et~al.} 2013,
  \aap, 558, A33, \dodoi{10.1051/0004-6361/201322068}

\bibitem[{{Astropy Collaboration} {et~al.}(2018){Astropy Collaboration},
  {Price-Whelan}, {Sip{\H{o}}cz}, {G{\"u}nther}, {Lim}, {Crawford}, {Conseil},
  {Shupe}, {Craig}, {Dencheva}, {Ginsburg}, {VanderPlas}, {Bradley},
  {P{\'e}rez-Su{\'a}rez}, {de Val-Borro}, {Aldcroft}, {Cruz}, {Robitaille},
  {Tollerud}, {Ardelean}, {Babej}, {Bach}, {Bachetti}, {Bakanov}, {Bamford},
  {Barentsen}, {Barmby}, {Baumbach}, {Berry}, {Biscani}, {Boquien}, {Bostroem},
  {Bouma}, {Brammer}, {Bray}, {Breytenbach}, {Buddelmeijer}, {Burke},
  {Calderone}, {Cano Rodr{\'\i}guez}, {Cara}, {Cardoso}, {Cheedella}, {Copin},
  {Corrales}, {Crichton}, {D'Avella}, {Deil}, {Depagne}, {Dietrich}, {Donath},
  {Droettboom}, {Earl}, {Erben}, {Fabbro}, {Ferreira}, {Finethy}, {Fox},
  {Garrison}, {Gibbons}, {Goldstein}, {Gommers}, {Greco}, {Greenfield},
  {Groener}, {Grollier}, {Hagen}, {Hirst}, {Homeier}, {Horton}, {Hosseinzadeh},
  {Hu}, {Hunkeler}, {Ivezi{\'c}}, {Jain}, {Jenness}, {Kanarek}, {Kendrew},
  {Kern}, {Kerzendorf}, {Khvalko}, {King}, {Kirkby}, {Kulkarni}, {Kumar},
  {Lee}, {Lenz}, {Littlefair}, {Ma}, {Macleod}, {Mastropietro}, {McCully},
  {Montagnac}, {Morris}, {Mueller}, {Mumford}, {Muna}, {Murphy}, {Nelson},
  {Nguyen}, {Ninan}, {N{\"o}the}, {Ogaz}, {Oh}, {Parejko}, {Parley}, {Pascual},
  {Patil}, {Patil}, {Plunkett}, {Prochaska}, {Rastogi}, {Reddy Janga},
  {Sabater}, {Sakurikar}, {Seifert}, {Sherbert}, {Sherwood-Taylor}, {Shih},
  {Sick}, {Silbiger}, {Singanamalla}, {Singer}, {Sladen}, {Sooley},
  {Sornarajah}, {Streicher}, {Teuben}, {Thomas}, {Tremblay}, {Turner},
  {Terr{\'o}n}, {van Kerkwijk}, {de la Vega}, {Watkins}, {Weaver}, {Whitmore},
  {Woillez}, {Zabalza}, \& {Astropy Contributors}}]{2018Astropy}
{Astropy Collaboration}, {Price-Whelan}, A.~M., {Sip{\H{o}}cz}, B.~M., {et~al.}
  2018, \aj, 156, 123, \dodoi{10.3847/1538-3881/aabc4f}

\bibitem[{{Baade} \& {Zwicky}(1934)}]{1934Baade}
{Baade}, W., \& {Zwicky}, F. 1934, Physical Review, 46, 76,
  \dodoi{10.1103/PhysRev.46.76.2}

\bibitem[{{Blandford} \& {Ostriker}(1978)}]{1978Blandford}
{Blandford}, R.~D., \& {Ostriker}, J.~P. 1978, \apjl, 221, L29,
  \dodoi{10.1086/182658}

\bibitem[{{Breitschwerdt} {et~al.}(1991){Breitschwerdt}, {McKenzie}, \&
  {Voelk}}]{1991Breitschwerdt}
{Breitschwerdt}, D., {McKenzie}, J.~F., \& {Voelk}, H.~J. 1991, \aap, 245, 79

\bibitem[{{Bustard} \& {Zweibel}(2021)}]{2021Bustard}
{Bustard}, C., \& {Zweibel}, E.~G. 2021, \apj, 913, 106,
  \dodoi{10.3847/1538-4357/abf64c}

\bibitem[{{Caprioli} \& {Spitkovsky}(2014)}]{2014Caprioli}
{Caprioli}, D., \& {Spitkovsky}, A. 2014, \apj, 783, 91,
  \dodoi{10.1088/0004-637X/783/2/91}

\bibitem[{{Chan} {et~al.}(2022){Chan}, {Kere{\v{s}}}, {Gurvich}, {Hopkins},
  {Trapp}, {Ji}, \& {Faucher-Gigu{\`e}re}}]{2022Chan}
{Chan}, T.~K., {Kere{\v{s}}}, D., {Gurvich}, A.~B., {et~al.} 2022, \mnras, 517,
  597, \dodoi{10.1093/mnras/stac2236}

\bibitem[{{Chan} {et~al.}(2019){Chan}, {Kere{\v{s}}}, {Hopkins}, {Quataert},
  {Su}, {Hayward}, \& {Faucher-Gigu{\`e}re}}]{2019Chan}
{Chan}, T.~K., {Kere{\v{s}}}, D., {Hopkins}, P.~F., {et~al.} 2019, \mnras, 488,
  3716, \dodoi{10.1093/mnras/stz1895}

\bibitem[{{Desiati} \& {Zweibel}(2014)}]{2014Desiati}
{Desiati}, P., \& {Zweibel}, E.~G. 2014, \apj, 791, 51,
  \dodoi{10.1088/0004-637X/791/1/51}

\bibitem[{{Diesing} \& {Caprioli}(2018)}]{2018Diesing}
{Diesing}, R., \& {Caprioli}, D. 2018, \prl, 121, 091101,
  \dodoi{10.1103/PhysRevLett.121.091101}

\bibitem[{{Draine}(2011)}]{2011Draine}
{Draine}, B.~T. 2011, {Physics of the Interstellar and Intergalactic Medium}

\bibitem[{{Elia} {et~al.}(2022){Elia}, {Molinari}, {Schisano}, {Soler},
  {Merello}, {Russeil}, {Veneziani}, {Zavagno}, {Noriega-Crespo}, {Olmi},
  {Benedettini}, {Hennebelle}, {Klessen}, {Leurini}, {Paladini}, {Pezzuto},
  {Traficante}, {Eden}, {Martin}, {Sormani}, {Coletta}, {Colman}, {Plume},
  {Maruccia}, {Mininni}, \& {Liu}}]{2022ApJ...941..162E}
{Elia}, D., {Molinari}, S., {Schisano}, E., {et~al.} 2022, \apj, 941, 162,
  \dodoi{10.3847/1538-4357/aca27d}

\bibitem[{{Evoli} {et~al.}(2019){Evoli}, {Aloisio}, \& {Blasi}}]{2019Evoli}
{Evoli}, C., {Aloisio}, R., \& {Blasi}, P. 2019, \prd, 99, 103023,
  \dodoi{10.1103/PhysRevD.99.103023}

\bibitem[{{Evoli} {et~al.}(2020){Evoli}, {Morlino}, {Blasi}, \&
  {Aloisio}}]{2020Evoli}
{Evoli}, C., {Morlino}, G., {Blasi}, P., \& {Aloisio}, R. 2020, \prd, 101,
  023013, \dodoi{10.1103/PhysRevD.101.023013}

\bibitem[{{Farber} {et~al.}(2018){Farber}, {Ruszkowski}, {Yang}, \&
  {Zweibel}}]{2018Farber}
{Farber}, R., {Ruszkowski}, M., {Yang}, H. Y.~K., \& {Zweibel}, E.~G. 2018,
  \apj, 856, 112, \dodoi{10.3847/1538-4357/aab26d}

\bibitem[{{Ferri{\`e}re}(2001)}]{2001Ferriere}
{Ferri{\`e}re}, K.~M. 2001, Reviews of Modern Physics, 73, 1031,
  \dodoi{10.1103/RevModPhys.73.1031}

\bibitem[{{Giz} \& {Shu}(1993)}]{1993Giz}
{Giz}, A.~T., \& {Shu}, F.~H. 1993, \apj, 404, 185, \dodoi{10.1086/172267}

\bibitem[{{Guo} \& {Oh}(2008)}]{2008Guo}
{Guo}, F., \& {Oh}, S.~P. 2008, \mnras, 384, 251,
  \dodoi{10.1111/j.1365-2966.2007.12692.x}

\bibitem[{{Habegger} {et~al.}(2024){Habegger}, {Ho}, {Yuen}, \&
  {Zweibel}}]{2024Habegger}
{Habegger}, R., {Ho}, K.~W., {Yuen}, K.~H., \& {Zweibel}, E.~G. 2024, \apj,
  974, 17, \dodoi{10.3847/1538-4357/ad67da}

\bibitem[{{Habegger} {et~al.}(2023){Habegger}, {Zweibel}, \&
  {Wong}}]{2023Habegger}
{Habegger}, R., {Zweibel}, E.~G., \& {Wong}, S. 2023, \apj, 951, 99,
  \dodoi{10.3847/1538-4357/accf8e}

\bibitem[{{Harris} {et~al.}(2020){Harris}, {Millman}, {van der Walt},
  {Gommers}, {Virtanen}, {Cournapeau}, {Wieser}, {Taylor}, {Berg}, {Smith},
  {Kern}, {Picus}, {Hoyer}, {van Kerkwijk}, {Brett}, {Haldane}, {del R{\'\i}o},
  {Wiebe}, {Peterson}, {G{\'e}rard-Marchant}, {Sheppard}, {Reddy}, {Weckesser},
  {Abbasi}, {Gohlke}, \& {Oliphant}}]{2020NumPy}
{Harris}, C.~R., {Millman}, K.~J., {van der Walt}, S.~J., {et~al.} 2020, \nat,
  585, 357, \dodoi{10.1038/s41586-020-2649-2}

\bibitem[{{Heintz} {et~al.}(2020){Heintz}, {Bustard}, \&
  {Zweibel}}]{2020Heintz}
{Heintz}, E., {Bustard}, C., \& {Zweibel}, E.~G. 2020, \apj, 891, 157,
  \dodoi{10.3847/1538-4357/ab7453}

\bibitem[{{Heintz} \& {Zweibel}(2018)}]{2018Heintz}
{Heintz}, E., \& {Zweibel}, E.~G. 2018, \apj, 860, 97,
  \dodoi{10.3847/1538-4357/aac208}

\bibitem[{{Ho} {et~al.}(2024){Ho}, {Yuen}, \& {Lazarian}}]{2024Ho}
{Ho}, K.~W., {Yuen}, K.~H., \& {Lazarian}, A. 2024, arXiv e-prints,
  arXiv:2407.14199, \dodoi{10.48550/arXiv.2407.14199}

\bibitem[{{Hosking} {et~al.}(2024){Hosking}, {Wasserman}, \&
  {Cowley}}]{2024Hosking}
{Hosking}, D.~N., {Wasserman}, D., \& {Cowley}, S.~C. 2024, arXiv e-prints,
  arXiv:2401.01336, \dodoi{10.48550/arXiv.2401.01336}

\bibitem[{{Hunter}(2007)}]{2007Matplotlib}
{Hunter}, J.~D. 2007, Computing in Science and Engineering, 9, 90,
  \dodoi{10.1109/MCSE.2007.55}

\bibitem[{{Inoue} {et~al.}(2006){Inoue}, {Inutsuka}, \& {Koyama}}]{2006Inoue}
{Inoue}, T., {Inutsuka}, S.-i., \& {Koyama}, H. 2006, \apj, 652, 1331,
  \dodoi{10.1086/508334}

\bibitem[{{Jiang} \& {Oh}(2018)}]{2018Jiang}
{Jiang}, Y.-F., \& {Oh}, S.~P. 2018, \apj, 854, 5,
  \dodoi{10.3847/1538-4357/aaa6ce}

\bibitem[{{Jones} {et~al.}(2001){Jones}, {Lukasiak}, {Ptuskin}, \&
  {Webber}}]{2001Jones}
{Jones}, F.~C., {Lukasiak}, A., {Ptuskin}, V., \& {Webber}, W. 2001, \apj, 547,
  264, \dodoi{10.1086/318358}

\bibitem[{{Kim} {et~al.}(2023){Kim}, {Kim}, {Gong}, \&
  {Ostriker}}]{2023ApJ...946....3K}
{Kim}, C.-G., {Kim}, J.-G., {Gong}, M., \& {Ostriker}, E.~C. 2023, \apj, 946,
  3, \dodoi{10.3847/1538-4357/acbd3a}

\bibitem[{{Kim} \& {Ostriker}(2015)}]{2015Kim}
{Kim}, C.-G., \& {Ostriker}, E.~C. 2015, \apj, 802, 99,
  \dodoi{10.1088/0004-637X/802/2/99}

\bibitem[{{Lacki} {et~al.}(2011){Lacki}, {Thompson}, {Quataert}, {Loeb}, \&
  {Waxman}}]{2011Lacki}
{Lacki}, B.~C., {Thompson}, T.~A., {Quataert}, E., {Loeb}, A., \& {Waxman}, E.
  2011, \apj, 734, 107, \dodoi{10.1088/0004-637X/734/2/107}

\bibitem[{{McCallum} {et~al.}(2025){McCallum}, {Wood}, {Benjamin},
  {Krishnarao}, \& {McLeod}}]{2025arXiv250617689M}
{McCallum}, L., {Wood}, K., {Benjamin}, R., {Krishnarao}, D., \& {McLeod},
  A.~F. 2025, arXiv e-prints, arXiv:2506.17689,
  \dodoi{10.48550/arXiv.2506.17689}

\bibitem[{{McClure-Griffiths} {et~al.}(2023){McClure-Griffiths},
  {Stanimirovi{\'c}}, \& {Rybarczyk}}]{2023McClureGriffiths}
{McClure-Griffiths}, N.~M., {Stanimirovi{\'c}}, S., \& {Rybarczyk}, D.~R. 2023,
  \araa, 61, 19, \dodoi{10.1146/annurev-astro-052920-104851}

\bibitem[{{McKenzie} \& {Voelk}(1982)}]{1982Mckenzie}
{McKenzie}, J.~F., \& {Voelk}, H.~J. 1982, \aap, 116, 191

\bibitem[{{Newcomb}(1961)}]{1961Newcomb}
{Newcomb}, W.~A. 1961, Physics of Fluids, 4, 391, \dodoi{10.1063/1.1706342}

\bibitem[{{O'Neill} {et~al.}(2024){O'Neill}, {Zucker}, {Goodman}, \&
  {Edenhofer}}]{2024ApJ...973..136O}
{O'Neill}, T.~J., {Zucker}, C., {Goodman}, A.~A., \& {Edenhofer}, G. 2024,
  \apj, 973, 136, \dodoi{10.3847/1538-4357/ad61de}

\bibitem[{{Owen} {et~al.}(2023){Owen}, {Wu}, {Inoue}, {Yang}, \&
  {Mitchell}}]{2023Owen}
{Owen}, E.~R., {Wu}, K., {Inoue}, Y., {Yang}, H. Y.~K., \& {Mitchell}, A. M.~W.
  2023, Galaxies, 11, 86, \dodoi{10.3390/galaxies11040086}

\bibitem[{{Parker}(1966)}]{1966Parker}
{Parker}, E.~N. 1966, \apj, 145, 811, \dodoi{10.1086/148828}

\bibitem[{{Rathjen} {et~al.}(2021){Rathjen}, {Naab}, {Girichidis}, {Walch},
  {W{\"u}nsch}, {Dinnbier}, {Seifried}, {Klessen}, \&
  {Glover}}]{2021MNRAS.504.1039R}
{Rathjen}, T.-E., {Naab}, T., {Girichidis}, P., {et~al.} 2021, \mnras, 504,
  1039, \dodoi{10.1093/mnras/stab900}

\bibitem[{{Rosen} \& {Bregman}(1995)}]{1995Rosen}
{Rosen}, A., \& {Bregman}, J.~N. 1995, \apj, 440, 634, \dodoi{10.1086/175303}

\bibitem[{{Ruszkowski} \& {Pfrommer}(2023)}]{2023Ruszkowski}
{Ruszkowski}, M., \& {Pfrommer}, C. 2023, \aapr, 31, 4,
  \dodoi{10.1007/s00159-023-00149-2}

\bibitem[{{Rybarczyk} {et~al.}(2024){Rybarczyk}, {Wenger}, \&
  {Stanimirovi{\'c}}}]{2024Rybarczyk}
{Rybarczyk}, D.~R., {Wenger}, T.~V., \& {Stanimirovi{\'c}}, S. 2024, \apj, 975,
  167, \dodoi{10.3847/1538-4357/ad79f7}

\bibitem[{{Schmidt}(1959)}]{1959ApJ...129..243S}
{Schmidt}, M. 1959, \apj, 129, 243, \dodoi{10.1086/146614}

\bibitem[{{Semenov} {et~al.}(2021){Semenov}, {Kravtsov}, \&
  {Caprioli}}]{2021Semenov}
{Semenov}, V.~A., {Kravtsov}, A.~V., \& {Caprioli}, D. 2021, \apj, 910, 126,
  \dodoi{10.3847/1538-4357/abe2a6}

\bibitem[{{Shimoda} \& {Inutsuka}(2022)}]{2022ApJ...926....8S}
{Shimoda}, J., \& {Inutsuka}, S.-i. 2022, \apj, 926, 8,
  \dodoi{10.3847/1538-4357/ac4110}

\bibitem[{Stone {et~al.}(2020)Stone, Tomida, White, \& Felker}]{2020Stone}
Stone, J.~M., Tomida, K., White, C.~J., \& Felker, K.~G. 2020, The
  Astrophysical Journal Supplement Series, 249, 4,
  \dodoi{10.3847/1538-4365/ab929b}

\bibitem[{{Tharakkal} {et~al.}(2023){Tharakkal}, {Shukurov}, {Gent}, {Sarson},
  {Snodin}, \& {Rodrigues}}]{2023Tharakkal}
{Tharakkal}, D., {Shukurov}, A., {Gent}, F.~A., {et~al.} 2023, \mnras, 525,
  5597, \dodoi{10.1093/mnras/stad2610}

\bibitem[{{Thomas} {et~al.}(2024){Thomas}, {Pfrommer}, \&
  {Pakmor}}]{2024Thomas}
{Thomas}, T., {Pfrommer}, C., \& {Pakmor}, R. 2024, arXiv e-prints,
  arXiv:2405.13121, \dodoi{10.48550/arXiv.2405.13121}

\bibitem[{{Townsend}(2009)}]{2009Townsend}
{Townsend}, R.~H.~D. 2009, \apjs, 181, 391, \dodoi{10.1088/0067-0049/181/2/391}

\bibitem[{{Tsung} {et~al.}(2023){Tsung}, {Oh}, \& {Bustard}}]{2023Tsung}
{Tsung}, T. H.~N., {Oh}, S.~P., \& {Bustard}, C. 2023, \mnras, 526, 3301,
  \dodoi{10.1093/mnras/stad2720}

\bibitem[{{Tsung} {et~al.}(2022){Tsung}, {Oh}, \& {Jiang}}]{2022Tsung}
{Tsung}, T. H.~N., {Oh}, S.~P., \& {Jiang}, Y.-F. 2022, \mnras, 513, 4464,
  \dodoi{10.1093/mnras/stac1123}

\bibitem[{{van der Walt} {et~al.}(2011){van der Walt}, {Colbert}, \&
  {Varoquaux}}]{2011NumPy}
{van der Walt}, S., {Colbert}, S.~C., \& {Varoquaux}, G. 2011, Computing in
  Science and Engineering, 13, 22, \dodoi{10.1109/MCSE.2011.37}

\bibitem[{{Walch} {et~al.}(2015){Walch}, {Girichidis}, {Naab}, {Gatto},
  {Glover}, {W{\"u}nsch}, {Klessen}, {Clark}, {Peters}, {Derigs}, \&
  {Baczynski}}]{2015MNRAS.454..238W}
{Walch}, S., {Girichidis}, P., {Naab}, T., {et~al.} 2015, \mnras, 454, 238,
  \dodoi{10.1093/mnras/stv1975}

\bibitem[{{Wiener} {et~al.}(2017){Wiener}, {Oh}, \&
  {Zweibel}}]{2017MNRAS.467..646W}
{Wiener}, J., {Oh}, S.~P., \& {Zweibel}, E.~G. 2017, \mnras, 467, 646,
  \dodoi{10.1093/mnras/stx109}

\bibitem[{{Wiener} {et~al.}(2019){Wiener}, {Zweibel}, \&
  {Ruszkowski}}]{2019Wiener}
{Wiener}, J., {Zweibel}, E.~G., \& {Ruszkowski}, M. 2019, \mnras, 489, 205,
  \dodoi{10.1093/mnras/stz2007}

\bibitem[{{Wolfram Research Inc.}(2023)}]{Mathematica}
{Wolfram Research Inc.} 2023, Mathematica, Version 13.3

\bibitem[{{Zweibel}(2017)}]{2017Zweibel}
{Zweibel}, E.~G. 2017, Physics of Plasmas, 24, 055402,
  \dodoi{10.1063/1.4984017}

\end{thebibliography}
\bibliographystyle{aasjournal}



\end{document}